\documentclass[useAMS,usenatbib,usegraphicx]{mn2e}

\title[A composition-dependent SPH model for chemical evolution and cooling]{Chemical evolution of galaxies. I. A composition-dependent SPH model for chemical evolution and cooling}

\author[F.J. Mart\'inez-Serrano, A. Serna, R. Dom\'inguez-Tenreiro and M. Moll\'a]{F.J.
Mart\'inez-Serrano$^{1}$, A. Serna$^{1}$, R. Dom\'inguez-Tenreiro$^{2}$ and M.
Moll\'a$^{3}$\\
$^{1}$Depto. de F\'isica y A.C., Universidad Miguel Hern\'andez,	
03206 Elche, Alicante, Spain\\
 $^{2}$Depto. de F\'isica Te\'orica, Universidad Aut\'onoma de Madrid,
28040 Cantoblanco, Madrid, Spain\\
$^{3}$Depto. de Investigaci\'on B\'asica, C.I.E.M.A.T., Avda. Complutense 22, 28040 Madrid, Spain}

\begin{document}

\date{\today}


\maketitle
\begin{abstract}
We describe an SPH model for chemical enrichment and radiative cooling in cosmological simulations of structure formation. This model includes: i) the delayed gas restitution from stars by means of a probabilistic approach designed to reduce the statistical noise and, hence, to allow for the study of the inner chemical structure of objects with moderately high numbers of particles; ii) the full dependence of metal production on the detailed chemical composition of stellar particles by using, for the first time in SPH codes, the Q$_{ij}$ matrix formalism that relates each nucleosynthetic product to its sources; and iii) the full dependence of radiative cooling on the detailed chemical composition of gas particles, achieved through a fast algorithm using a new metallicity parameter $\zeta(T)$ that gives the weight of each element on the total cooling function. The resolution effects and the results obtained from this SPH chemical model have been tested by comparing its predictions in different problems with known theoretical solutions. We also present some preliminary results on the chemical properties of elliptical galaxies found in self-consistent cosmological simulations. Such simulations show that the above $\zeta$-cooling method is important to prevent an overestimation of the metallicity-dependent cooling rate, whereas the Q$_{ij}$ formalism is important to prevent a significant underestimation of the [$\alpha$/Fe] ratio in simulated galaxy-like objects.
\end{abstract}

\label{firstpage}

\section{Introduction}
\label{introduction}

Chemical abundances are a rich reservoir of information which can help uncover how the formation of galaxies took place. Indeed, even after epochs where much information is lost through phase mixing or violent relaxation, stars still retain important clues about the evolutionary histories of the objects from which they come \citep[e.g.,][]{2002ARA&A..40..487F}. The earliest studies of the chemical evolution of galaxies were based on the well-known Closed Box model \citep{1975VA.....19..299L,1980FCPh....5..287T}, that played an important role in obtaining a first insight into this problem. This model considers galaxies as one-zone systems with a constant total mass and instantaneous recycling and mixing of matter. Because of such simplifying assumptions, the Closed Box model provides no idea about the internal structure of galaxies and leads to some predictions that are inconsistent with the distribution of G-dwarfs in the solar vicinity, as well as with other observations \citep{1975MNRAS.172...13P,2002ApJ...566..252V}. 

After the seminal work of \citet{1983MNRAS.204..791L,1985ApJ...290..154L}, other approaches \citep[e.g.,][]{1986A&A...158...60D,1987A&A...185...51M,1987ApJ...315..451C,1989MNRAS.238..133S,1989MNRAS.239..885M,1994ApJ...427..745F,1997ApJ...477..765C} have considered multi-zone chemical evolution models with external gas infall. Numerical computations based on these kinds of models have been successful to describe the radial distribution of abundances in our Galaxy \citep{1994ApJ...427..745F,1999MNRAS.307..857B} and in other spiral disks \citep{1996ApJ...466..668M,2000MNRAS.312..398B,2005MNRAS.358..521M}. The success of most of these models is based on appropriate spatial variations of the ratio between the star formation rate and gas infall rate, which are achieved by using different code input parameters for different galaxy types. It is then important to compute more realistic models where such rates naturally appear, within a cosmological context, as a result of the physical processes involved in the formation and dynamical evolution of galaxies. Due to the complexity of this problem, it must be addressed from approaches like, e.g., hydrodynamic N-body simulations.

Among the different methods developed for the modelling of complex hydrodynamic phenomena, the Smooth Particle Hydrodynamics (SPH) technique \citep{1992ARA&A..30..543M} is one of the most widely used in astrophysics. Due to its Lagrangian nature, the evolving distribution of the gas, dark matter and stellar components can be easily followed in a self-consistent way. The first approach to include chemical evolution in an SPH code was proposed by \citet{1994A&A...281L..97S,1995MNRAS.276..549S}, followed by \citet{1996A&A...315..105R}, \citet{1998MNRAS.297.1021C}, \citet{1999A&A...348..371B}, \citet{2003MNRAS.340..908K} and \citet{2004MNRAS.347..740K}, among others. All these works focussed on the chemical enrichment of isolated objects, or groups of objects, formed from pre-prepared initial conditions. Recent works \citep[see, e.g.,][]{2001MNRAS.325...34M,2005MNRAS.364..552S,2007MNRAS.376.1465K} have extended SPH simulations to study the detailed chemical enrichment of galaxies within a full cosmological context, where mergers and interactions have important effects.

A new degree of sophistication in the SPH-modelling of chemical enrichment was introduced by \citet[][hereafter LPC02]{2002MNRAS.330..821L}, who proposed a stochastic algorithm to completely remove the assumption of instantaneous recycling of stellar ejecta. Such a delayed gas restitution from stars has non-negligible dynamical effects on N-body disks, as discussed by \citet{2001A&A...376...85J,2004Ap&SS.289..441J}, and a direct impact on the resulting abundance gradients. The algorithm of LPC02 takes into account the non-instantaneous gas restitution from stars through a method where the total number of baryonic particles remains constant and no dynamically hybrid particles are present: they are either fully collisional or fully collisionless. The presence of such hybrid particles would introduce spurious dynamical effects like, e.g., stars that follow for a while the evolution of the gas. The LPC02 approach, already used in some studies \citep{2005MNRAS.357..478S,2005MNRAS.361..983R,2006MNRAS.371..548R}, has the advantage that it leads to objects with reliable {\it average} values for their chemical properties. Nevertheless, due to its statistical nature, its main disadvantage is that large numbers of particles per object are needed to avoid an excessively high scatter around the average values. 

Together with the inclusion of delayed gas restitution from stars, a necessary further improvement on the SPH modelling of chemical enrichment is the full dependence of both metal production and radiative cooling on the detailed chemical composition of star and gas particles, respectively. Indeed, the metal production of stars is usually included in SPH codes on the base of up-to-date libraries with stellar models and their corresponding ejecta for stars with different masses and metallicities. Such libraries \citep[e.g.,][]{1998A&A...334..505P,2005A&A...432..861G} generally assume solar relative abundances of the various species for any total metallicity $Z$. Nevertheless, it is well known that within a given $Z$ different chemical compositions are possible. Chemical evolution models show in fact that abundance ratios are not constant in the course of galactic evolution \citep[see, e.g.,][]{1998A&A...334..505P}. Deviations from solar proportions, particularly the ratios of $\alpha$-peak to Fe-peak elements, are also observed in different stellar systems. Supra-solar values for these ratios are observed in metal-poor stars of our Galaxy as well as in globular clusters \citep{2002A&A...395..761K,2002AJ....124..828L}, where [$\alpha$/Fe] seems to increase with decreasing metallicity \citep{2005A&A...439..997P}, reaching values that vary from some $\alpha$ elements to others: Mg seems to converge towards [Mg/Fe] $\sim$ 0.4 for very metal-poor stars \citep[see][]{1999Ap&SS.265..265F} whereas oxygen trends towards [O/Fe] $\sim 0.6$ \citep[see][]{1999Ap&SS.265..171R}. Such an $\alpha$-enhancement is also observed in spiral galaxy bulges \citep{1994MNRAS.271...39S} as well as in most of the central regions of elliptical galaxies \citep{1984ApJ...287..586B,1988ApJ...328..440B,1992ApJ...398...69W,1993MNRAS.265..731G,1993MNRAS.265..553C,1994AJ....107..946D,
1994ApJS...95..107W}, where the [$\alpha$/Fe] ratio increases with velocity dispersion and hence with mass. Since the $\alpha$-enhancement is commonly interpreted as the result of large and short star formation events at early times, it could provide us with a very strong constraint for models of galaxy formation. To accurately follow the evolution of this or any other abundance ratio, the assumption of solar proportions must be relaxed in cosmological simulations when computing the production of the different chemical elements. 

On the other hand, the chemical content of the gas in the interstellar medium (ISM) is enriched by metal ejecta of different stars and then mixed through complex processes involving both local diffusion and motions at larger scales. Consequently, the gas of cosmological simulations (as well as in real objects) has a non-trivial mixture of metals with trends that are not necessarily the same as in the solar vicinity. Since the radiative cooling rate of gas closely depends on its metal composition, a realistic modelling of galaxy formation must consider a metal-dependent cooling function. As pointed out by some authors \citep[e.g.,][]{2006MNRAS.371..548R}, ideally one would need cooling functions for different relative proportions of different elements. Nevertheless, the lack of a fast algorithm to implement such a composition-dependent cooling rate has forced the use in cosmological simulations of the \citet{1993ApJS...88..253S} functions, given for chemical compositions mimicking those in the solar vicinity (scaled, according to the total metallicity, from primordial to solar abundances).

In this paper we present a new SPH-model for chemical evolution and cooling. Our model includes: i) the delayed gas restitution from stars by means of a probabilistic approach based on the LPC02 scheme. Such a scheme has been modified to reduce the statistical noise and, hence, to allow for the study of the inner chemical structure of objects with moderately high numbers of particles; ii) the full dependence of metal production on the detailed chemical composition of stellar particles. To this end, the chemical production of a stellar particle is computed by using the Q$_{ij}$ matrix formalism \citep{1973ApJ...186...51T}, that relates each nucleosynthetic product to its sources; and iii) the full dependence of radiative cooling on the detailed chemical composition of gas particles. This latter issue is achieved through a fast algorithm where the cooling rate is not computed by using the total metallicity as the scaling parameter. We use instead a parameter $\zeta(T)$ defined as a linear combination of the abundance of different chemical species. The coefficients of such a linear combination depend on temperature and give the weight of each element on the total cooling function.

The work presented in this paper will focus on the description and testing of our chemical model. In order to analyse whether this model is able to produce galactic objects with chemical properties consistent with observations, we also report some first results on the cosmological formation of elliptical galaxies. Such simulations have been carried out by using the DEVA code \citep{2003ApJ...597..878S}. Nevertheless, the model presented in this paper is not limited to any particular code. In forthcoming papers we will analyse for larger galaxy samples some important chemical properties like, e.g., the metallicity distribution functions (MDF) and abundance gradients within individual objects, as well as the mass (or luminosity) relation with age, metallicity and [$\alpha$/Fe].

This paper is organised as follows. In Section 2 we describe our model for chemical evolution, whereas Section 3 presents our algorithm to compute composition-dependent cooling functions. Some synthetic tests as well as first results on cosmological simulations are shown in Section 4. Finally, in Section 5 we summarise our main results.

\section{The Chemical Evolution Model}
\label{sec:CEM}

As in any particle-based scheme, both the dark matter and baryonic content of the simulation box are sampled in our model by using a discrete number $N$ of particles. Each baryonic particle can be either in the form of gas or in the form of a stellar particle representing a single stellar population (SSP) of mass $m$ and age $t$. We do not consider hybrid particles simultaneously hosting both a stellar and gas content. Star formation (SF) and gas restitution from stars are then modelled by turning gas into stars, or stars into gas, respectively. 

Within the above scheme the chemical evolution model must provide us with procedures to compute: i) when and how gas particles are turned into stars (star formation); ii) when and how stellar particles are turned into gas (gas restitution); iii) the metal production of each stellar particle over a timestep $\Delta t$ (metal production); and iv) how the chemical production of stars is released and mixed through the gas component (metal ejection and diffusion).
We will now address each of these issues separately.

\subsection{Star Formation}

Star formation is commonly included in most SPH codes for galaxy formation. Different criteria have been proposed in the literature to decide when and how gas particles are turned into stars \citep[e.g.,][]{1992ApJ...391..502K,1994A&A...281L..97S,1997MNRAS.284..235Y,2001KFNT...17..213B,2003MNRAS.339..289S,2007A&A...473..733M}. In our current implementation of SF we consider that gas particles are eligible to form stars if they are located in a region with a convergent flow and a gas density higher than a given threshold, $\rho_{th}$. In that case they are turned into stellar particles according to a Kennicutt--Schmidt law-like transformation rule \citep[see][]{1998ApJ...498..541K,2001MNRAS.324..313S},
\begin{equation}
\frac{d\rho_g}{dt}=-\frac{d\rho_{\ast}}{dt}=-\frac{c_{\ast}\rho_g}{t_g}\;,
\label{star-rate}
\end{equation}
where $\rho_g$ and $\rho_{\ast}$ are gas and stellar density, respectively, $c_{\ast}$ is a dimensionless star-formation efficiency parameter, and $t_g$ is a characteristic time-scale chosen to be equal to the maximum of the local gas-dynamical time $t_{dyn}=(4\pi G\rho_g)^{-1/2}$, and the local cooling time, $t_{cool} = u_i/\dot{u}_i$, where $u_i$ is the internal energy. Eq. (\ref{star-rate}) implies that the probability $p$ that a gas particle forms stars in a time $\Delta t$ is
\begin{equation}
p=1-e^{-c_{\ast}\Delta t/t_g}\;.
\end{equation}

As usual, we compute $p$ at each time step for all eligible gas particles and draw random numbers to decide which of them form stars in the time interval $[t,t + \Delta t]$. Then, each of these randomly selected gas particles is transformed into one stellar particle.

\subsection{Gas restitution}\label{sec:GR}

Once stars are present they return to the ISM part of their mass in form of
chemically processed gas. As already mentioned, each stellar particle is treated as an SSP of total mass $m$ and age $t$. Within each SSP, stellar masses are distributed according to a given Initial Mass Function (IMF), $\Phi(M)$. Throughout this paper we use the \citet{2003PASP..115..763C} IMF with a mass range of $[M_l,\, M_u]=[0.1,\, 100]$ M$_\odot$. Such an IMF is similar to other possible choices \citep[e.g.,][]{1955ApJ...121..161S,1998MNRAS.298..231K}, but provides a better fit to counts of low-mass stars \citep{2003ApJ...586L.133C,2003ApJS..149..289B}.

Individual stars of mass $M$ are characterised by a mean-lifetime $\tau(M)$. Therefore, within an SSP of age $t$, any star with $\tau(M)<t$ has already died so that part of its mass remains as a stellar remnant, $M_r$, while the rest should be in the form of gas ejected back to the ISM. The gas mass fraction, $E(t)$, of an SSP of age $t$ is then given by:
\begin{equation}\label{ET}
 E(t)=\int_{M(t)}^{M_u} \frac{M-M_r(M)}{M}\Phi(M)\,dM\;,
\end{equation}
where $M_u$ is the upper mass limit for stars in the IMF, and $M(t)$ is the mass
of stars with lifetime $t$ so that any individual star with $M>M(t)$ has already ejected a mass $M-M_r(M)$ of gas. We compute (\ref{ET}) by using the stellar mean-lifetimes $\tau(M)$ from the Geneva evolutionary tracks \citep{1992A&AS...96..269S} and the remnant masses $M_r(M)$ of \citet{2005A&A...432..861G} for $M\leq 8 M_{\odot}$, or \citet{1995ApJS..101..181W} for $M\ge 8 M_{\odot}$.

In a stochastic approach the above mass fraction of gas can be used to compute, at each timestep $\Delta t$, the probability that an entire stellar particle turns back into gas through a Monte Carlo method similar to that used for SF. Such a probability is given by (LPC02):
\begin{equation}
 p_g=\frac{E(t+\Delta t)-E(t)}{1-E(t)}=\frac{\int_t^{t + \Delta t} e(t')\,dt'}{1-E(t)}\;,
 \label{eq:probability}
\end{equation}
where
\begin{equation}
 e(t)=\frac{dE(t)}{dt}
\end{equation}
is the instantaneous rate of total ejecta, while $[1-E(t)]^{-1}$ is a correction factor that accounts for the stellar particles that have already turned to gas, so that the probability $p_g$ is not computed for them anymore.

\subsection{Metal production}\label{sec:metalprod}

We assume that initially all gas particles have a primordial composition. As part of stellar evolution, newly
produced elements are released to the surrounding ISM, either as stellar winds
or byproducts of supernovae (SN\ae) explosions. We consider the evolution of
the following elements: H, $^4$He, $^{12}$C, $^{13}$C, $^{14}$N, $^{16}$O,
$^{20}$Ne, $^{24}$Mg, $^{28}$Si, $^{32}$S, $^{40}$Ca and $^{56}$Fe.

For an SSP of age $t$, the instantaneous ejection rate of a given element $i$ can be expressed as:
\begin{equation}\label{eq:ei}
e_i(t)=p_i(t)+e(t)X_i\;,
\end{equation}
where $p_i(t)$ is the ejection rate of the newly synthesised element $i$ (i.e., the yield of $i$), while $e(t)X_i$ is the ejection rate of the element $i$ already present when the SSP was formed from a gas particle with abundance $X_i$. All these rates are expressed as quantities per total mass unit. 

As already quoted in $\S$ \ref{introduction}, the yield of any element $i$ depends on the detailed abundances of all other species $j$. In order to take into account such a dependence we use the Q$_{ij}$ formalism proposed by \citet{1973ApJ...186...51T}. Such a formalism links any ejected species to all its different nucleosynthetic sources, allowing the model to scale the ejecta with respect to the detailed initial composition of a star.

For an individual star of mass $M$, the mass fraction initially in the form of chemical species $j$, transformed and ejected as chemical species $i$, is written in the Q$_{ij}$ formalism as:

\begin{equation}
 Q_{ij}(M) = \frac{M_{ij,exp}}{M_{j}}=\frac{M_{ij,exp}}{X_{j}M}\;,
\nonumber
\end{equation}
where $X_{j}$ is the initial abundance of $j$, whereas $M_{ij,exp}$ is
the mass of $i$ that was synthesised from $j$ and finally expelled. 

For a whole SSP, the contribution to $e_i(t)$ due to $j$ is then given by
\begin{equation}
 q_{ij}(t)=-Q_{ij}(M(t))\Phi(M(t))\frac{\mathrm{d}M}{\mathrm{d}t}\;,
 \nonumber
\end{equation}
and, therefore, the total ejection rate of the chemical species $i$ can be written as:
\begin{equation} \label{eq:ejection}
 e_i(t)=\sum_j q_{ij}(t)X_j
\end{equation}
and the yield of $i$ can be obtained from Eqs. \ref{eq:ei} and \ref{eq:ejection} as
\begin{equation}
 p_i(t)=e_i(t)-e(t)X_i\;.
 \label{eq:yields}
\end{equation}

To compute $q_{ij}(t)$ we consider the element production from:

\begin{enumerate}

 \item Enrichment from low and intermediate mass (LIM) stars. We assume that all stars with masses in the range 0.8 - 8 M$_\odot$ end their life, after an AGB or TP-AGB phase, by the loss of their envelope that is ejected to the ISM as enriched gas in the form of planetary nebulae. We
use the set of stellar yields by \citet{2005A&A...432..861G}, which
include the effects of the third dredge-up for the TP-AGB stage and of the hot-bottom burning processes. In this set the primary nitrogen contribution of the LIM stars is higher for low metallicities \citep{2006A&A...450..509G} and, simultaneously, the total nitrogen yield is lower than in
other stellar yield sets \citep{1997A&AS..123..305V,2001A&A...370..194M}.
Models using these stellar yields produce results in excellent agreement with
observation of nitrogen from our Galaxy (halo and disk stars), from extragalactic HII regions, and from Damped-Lyman-Alpha galaxies \citep{2006A&A...450..509G,2006MNRAS.372.1069M}.

 \item Type II supernovae (SNII). We assume that stars more massive than 8
M$_\odot$ produce SNII. For the chemical production we adopt the yields of
\citet{1995ApJS..101..181W}\footnote{The Fe ejecta has been divided by 2,
following \citet{1995ApJS...98..617T}}. These yields include the elements
produced in pre-supernova evolution and SNII explosions for metallicities
between $Z=0$ and $Z_\odot$ and masses between 11 and 40 M$_\odot$\footnote{We
extrapolate linearly for ejecta up to the upper mass limit considered (100
M$\odot$).}. Unlike other sets \citep[e.g.,][]{1998A&A...334..505P}, these yields
do not consider the contribution of stellar winds. As shown by
\citet{2005A&A...432..861G,2006A&A...450..509G}, when stars suffer mass loss from winds they eject large quantities of helium and carbon from their envelopes.
This hinders the creation of oxygen and other heavier elements, and results in a
low yield of oxygen and a high yield of carbon.

 \item Type Ia supernovae (SNIa). We use the chemical yields from the W7 model
by \citet{1999ApJS..125..439I}, tabulated for two different metallicities
(solar and sub-solar). The SNIa rates are computed according to
\citet{2000MmSAI..71..435R} who provided us with a numerical table (private
communication) with the time evolution of the supernova rates for a single
stellar population. These rates were derived for several combinations of
possible candidates of binary system or SNIa (double degenerate, single
degenerate, etc.), considering all parameters (secondary lifetimes, orbital
velocities, distances between both stellar components, etc.) that determine the
conversion of a binary system into an SNIa explosion.
\end{enumerate}

All the above sets of stellar yields are used as input data to construct the
$q_{ij}(t)$ matrix according to the \citet{1992ApJ...387..138F} algorithm as
updated in \citet{2005A&A...432..861G}. Such an updated algorithm is based on
\citet{1995ApJ...443..536G} and \citet{1998A&A...334..505P}.

\subsection{Metal ejection and diffusion}

\subsubsection{Metal ejection}

How the chemical elements are distributed and mixed in the ISM is a complex problem that takes place at scales much smaller than the resolution reached in cosmological simulations. Most SPH implementations of chemical enrichment do not consider gas restitution from stars. Therefore, at each timestep $\Delta t$, the metals produced by a stellar particle are distributed through the neighbouring gas by means of, e.g., SPH-spreading. This simple procedure fails however when the gas restitution from stars is considered. Indeed, the most straightforward way of restoring gas is to perform the opposite process of star formation, and turn whole stellar particles into gas. However, when a stellar particle of age $t$ turns back into a gas particle, its metal production from stars with lifetimes longer than $t$ has not been considered yet. If, at that timestep, such a remaining metal production is computed and distributed over the neighbouring gas, particles with an extremely high metallicity could result. Otherwise (i.e., if the remaining metal production is ignored), simulations including gas restitution would systematically underestimate the chemical enrichment. A straightforward procedure to circumvent this problem consists in introducing new gas particles and/or modifying the individual masses of particles to progressively account for the gas restitution from stars. 

Within a scheme where both the individual mass and the total number of particles remain constant, as we consider here, the metal release from stars can be also addressed through a statistical approach like that proposed by LPC02. That is to say, we must evaluate the overall metal production of a large enough number $N$ of stellar particles with similar ages and initial metallicities, so that they themselves constitute an SSP. The way in which individual particles release their metals must then ensure the right overall chemical evolution for the whole set of $N$ particles. For example, a possible procedure suggested by LPC02 is based on the fact that, within a timestep $\Delta t$, a stellar particle of age $t$ theoretically ejects an amount $\int_t^{t + \Delta t}e(t')\,dt'$ of gas with chemical abundances given by
\begin{equation}
 X_i'(t)=\frac{\int_t^{t + \Delta t} e_i(t')\,d't}{\int_t^{t + \Delta t}
e(t')\,dt'}\;.
 \label{eq:newabund}
\end{equation}

\begin{figure*}
\begin{center}
\includegraphics[angle=-90,width=\textwidth]{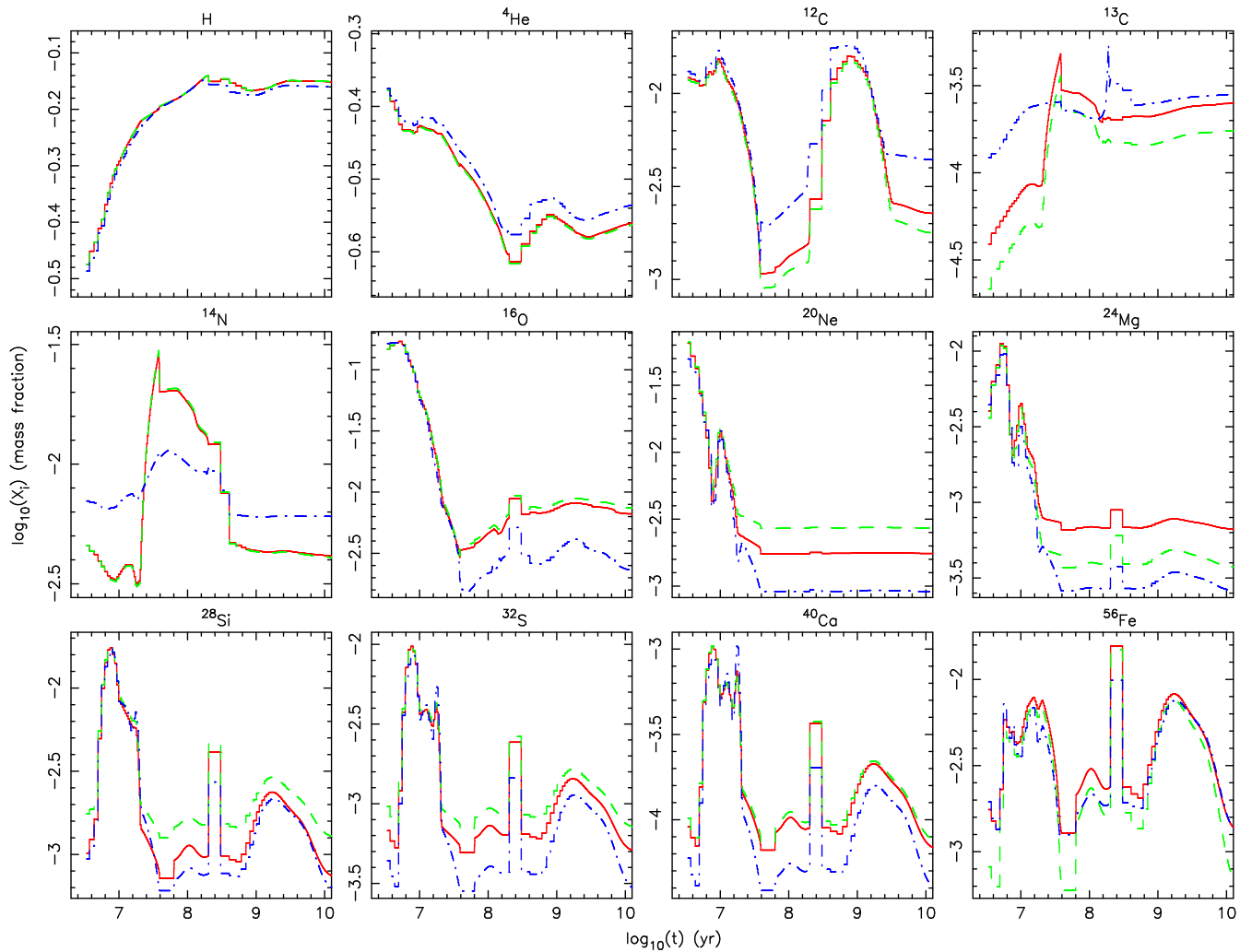}
\caption{\label{fig:yields} Composition of the ejecta as a function of time for three SSPs with solar total metallicity but different abundance ratios. The solid line corresponds to solar proportions, the dashed line to [O/Fe]=0.5, and the dashed-dotted line to [O/Fe]=-0.5. It can be seen how, even at identical total metallicity, in many cases the production of a particular isotope varies significantly. This effect is due solely to the use of the Q$_{ij}$ matrix formalism, since the stellar evolution models used in this work only take into account the initial total metallicity, which is the same for the three compositions.}
\end{center}
\end{figure*}

This procedure then considers that, as long as a particle remains as a stellar one, there is no gas restitution and metals are not released to the surrounding gas. On the contrary, if such a particle turns back into gas at some timestep $\Delta t$, the chemical content of the resulting gas-again particle is updated so that it corresponds to the composition of the gas released at such timestep, given by Eq. (\ref{eq:newabund}). Over a sufficiently large number $N$ of particles, constituting a whole SSP, this approach gives a fair representation of the overall metal production. The main limit of this method is that newly produced metals for a whole set of stellar particles are sampled by the few particles of different ages that turn back into gas within a given timestep, i.e., within a given cosmic age interval, $[t, t + \Delta t]$. Since the chemical composition of the gas released from an SSP is a function that strongly varies with time (see Fig. \ref{fig:yields}), a large number of particles is then needed to obtain a meaningful sampling, but also to avoid strong statistical noise which could lead to excessively high scatter in the resulting distribution of abundances.

In order to reduce the above statistical noise we have implemented a different procedure, in such a way that the metals produced by any stellar particle at a given timestep $[t,t + \Delta t]$ are distributed through the neighbouring gas at this timestep (this procedure is similar to that used in previous SPH implementations of chemical enrichment without restitution). Actually, since we incorporate the diffusion of metals through the gas (see below), just the nearest gas neighbour receives the yields of a stellar particle during each timestep. Note that, since stars do not follow the motion of gas particles, the identity of such nearest gas neighbour can change from one step to another. The composition $X_i'$ of the nearest gas neighbour\footnote{When a stellar particle is turned into gas, the resulting gas-again particle is considered as the nearest gas neighbour in that timestep.} of a stellar particle of age $t$ is updated according to:

\begin{equation}
 \Delta X_i' (t) =\frac{\int_t^{t + \Delta t} p_i(t')\,dt'}{1-E(t)}\;.
 \label{eq:spread}
\end{equation}

Here the yield $p_i(t)$ is given by (\ref{eq:yields}) and the factor $[1-E(t)]^{-1}$ converts yields per total mass unit to yields per stellar mass unit and then corrects for the missing metal production of stellar particles that were already transformed into gas. To see that is indeed the case, consider $N_0$ particles of mass $m$ that become stellar particles at $t_0=0$. Such $N_0$ particles can be regarded as an SSP. For each chemical species $i$, the 
mass production by this SSP during the time interval $[t, t+\Delta t]$ is:

\begin{equation}
\Delta M_i(t) = m N(t) \Delta X_i'(t)\;,
\end{equation}
where $m N(t) = m N_0 [1-E(t)]$ is the actual stellar mass available at time
$t$, after a fraction $E(t)$ of the stellar mass initially available at $t=0$
has been transformed again into gas. Taking into account (\ref{eq:spread}),
we can write:

\begin{equation}
\Delta M_i(t) = m N_0 \int_{t}^{t+\Delta t} p_i(t')dt'\;,\label{yield-SSP}
\end{equation}
and this is just the expression for the theoretical mass production
during the time interval $[t, t+\Delta t]$ of the $i$-th element by the whole stellar mass in the SSP, $m N_0$.

Note also that, since the yield definition implies $\sum_i p_i(t)=0$, Eq. (\ref{yield-SSP}) gives:

\begin{equation}
 \sum_i \Delta M_i(t) = 0\;\label{mass_ex}
\end{equation}

Therefore, when the composition of a gas particle is updated using (\ref{eq:spread}), there is no mass exchange between such a gas particle and the neighbouring stellar particle that releases its metals. The mass of each particle remains then unaltered and the only mechanism transforming stellar mass into gaseous mass is still that described in $\S$ \ref{sec:GR} for gas restitution.

\subsubsection{Metal diffusion in the gas}
\label{sec:diffusion}

Once the stellar production of metals has been released to the ISM, their redistribution through the gas component is governed by the turbulent motions of the gas. The turbulent mixing that takes place at subresolution scales can be properly modelled by a diffusion law at resolved scales \citep{taylor:21,2003PhRvE..67d6311K}.
\begin{equation}
\frac{\partial X_i}{\partial t} = D \vec{\nabla}^2 X_i\;.
\label{eqn:diff}
\end{equation}

The SPH formulation of the diffusion equation for a compressible
fluid has been given by \citet{2005RPPh...68.1703M}:

\begin{equation}
\frac{dX_i^a}{dt} = \sum_b K^{ab} (X_i^b - X_i^a)\;,
\label{eqn:diffsph}
\end{equation}
with
\begin{equation}
K^{ab} = \frac{m_b}{\rho_a \rho_b} \frac{4 D_a D_b}{D_a + D_b}
\frac{\left|\nabla_a W_{ab}\right|}{\left|\mathbf{r}_{ab}\right|}
\label{eqn:diffsph2}
\end{equation}
where the subindexes $a$ and $b$ are used to denote different SPH particles. As a first approximation, in this paper we have considered a constant diffusion parameter $D_a = D_b = D$ for all the gas particles. Except in the first three synthetic tests of $\S$ \ref{sec:synthetic}, where the instantaneous mixing of metals through the gas component is assumed, and in the diffusion test of $\S$ \ref{sec:testdif}, we have adopted the value $D=9.25\times 10^{26}$ cm$^2$/s as suggested by LPC02.

In our code, the diffusion equation (\ref{eqn:diffsph}) is computed at each timestep by considering all the active gas particles $a$, and their active neighbours $b$\footnote{Only active gas particles are considered in order to maintain the symmetry and ensure the abundance conservation in the diffusion process}. By active particles at timestep $[t,t + \Delta t]$, we mean those needing the update of their physical properties at that timestep. In order to solve Eq. (\ref{eqn:diffsph}) it must be noted that metal diffusion would introduce a new time-scale $(X/\dot{X})_{\textrm{diff}}$ and therefore, an additional criterion limiting the time stepping. Just like for the radiative cooling \citep[see][]{2003ApJ...597..878S}, such an additional $\Delta t$-control criterion can be circunvented by solving Eq. (\ref{eqn:diffsph}) in integrated form. We have then used the fact that, because of the Courant condition, the density field (and hence $K^{ab}$) is nearly constant over a timestep. Therefore, Eq. (\ref{eqn:diffsph}) can be anallytically integrated to find:
\begin{equation}
 \Delta X_i^a=X_i^a(t+\Delta t)-X_i^a(t)=\sum_b \Delta X_i^{ab}
\end{equation}
with
\begin{equation}
 \Delta X_i^{ab}=-\Delta X_i^{ba}=\frac{1}{2}\left(1-e^{-2 K^{ab} \Delta t} \right)(X_i^b(t) - X_i^a(t))\;.
\end{equation}
Such analytical expressions are those actually used in our implementation of the diffusion process.

\section{The composition-dependent cooling function}
\label{sec:cooling}

\subsection{The cooling model}

The publicly available MAPPINGS III code \citep[see][hereafter SD93, for a much
wider description]{1993ApJS...88..253S} consists of a detailed
cooling model for low and high temperatures, which includes
calculations for up to 16 atoms (H, He, C, N, O, Ne, Na, Mg, Al, Si,
S, Cl, Ar, Ca, Fe and Ni) and all stages of ionization. In this
code, the net cooling function of an optically-thin astrophysical
plasma is obtained by adding the contribution of different
processes: collisional line radiation (including fine-structure,
inter-system, and forbidden emission), $\Lambda_{\rm{lines}}$,
free-free and two-photon continuum radiation, $\Lambda_{\rm{cont}}$,
recombination processes with both cooling and heating effects,
$\Lambda_{\rm{rec}}$, photoionization heating, $\Lambda_{\rm{photo}}$, and
collisional ionization, $\Lambda_{\rm{coll}}$.
\begin{equation}\label{LambdaNet}
\Lambda_{\rm{net}}=\Lambda_{\rm{lines}}+\Lambda_{\rm{cont}}\pm
\Lambda_{\rm{rec}}-\Lambda_{\rm{photo}}+\Lambda_{\rm{coll}}
\end{equation}

All the detailed cooling computations in this paper have been
carried out by using MAPPINGS III. To obtain results easily
comparable to those reported by SD93, widely
used in cosmological simulations, we selected the algorithms and
assumptions of its previous version. In addition, all these
computations considered Collisional Ionization Equilibrium (CIE)
conditions. Therefore, the photoionization heating is insignificant,
leaving only collisional line radiation, continuum radiation, and
recombination heating as significant terms.

\subsection{Metallicity dependent cooling in cosmological simulations}

In principle, the accurate value of the normalised cooling function\footnote{Please note that we define the normalised cooling rate as $\Lambda_N=\Lambda_{\textrm{net}}/\rho^2$, instead of $\Lambda_N=\Lambda_{\textrm{net}}/n_e n_t$, where $n_e$ is the electron number density and $n_t$ the total ion number density. The resulting units for $\Lambda_N$ are erg cm$^3$ s$^{-1}$ g$^{-2}$. This approach eases the computation of $\Lambda_{\textrm{net}}$ in SPH codes.}
depends on both the local temperature and the detailed chemical composition:
\begin{equation}\label{CoolingDependence}
\Lambda_N=\Lambda_N(T,\mathbf{Z})
\end{equation}
where
\begin{equation}\label{eq:Z}
\mathbf{Z}=(Z_1, Z_2,..., Z_N)
\end{equation}
is a vector containing the abundance $Z_i$ of the $N=16$ chemical
species taken into account by MAPPINGS for the cooling rate. In a cosmological
simulation, one must deal with a huge number of gas particles with
different metal mixtures. A full computation (e.g., by directly
using MAPPINGS) of the cooling function of each particle according
to its detailed chemical composition would be very expensive in
terms of computing time.

To deal with the above difficulty, the most common approach consists
of reducing the $\mathbf{Z}$ dependence of the cooling function into a
much smaller number of parameters or, ideally, into just one
parameter $\zeta$.
\begin{equation}\label{CoolingRedux}
\Lambda_N=\Lambda_N(T,\zeta)
\end{equation}

For example, SD93 have considered the total metallicity, $Z_{\rm{tot}}$, as
their choice of $\zeta$. More specifically, the cooling function is computed and
tabulated by using chemical proportions interpolated, according to the total
metallicity $Z_{\rm{tot}}$, between primordial and solar abundances.
Such an approach allows for a fast implementation of a metallicity
dependent cooling function in cosmological simulations.

\begin{figure}
 \centering
 \includegraphics[width=\columnwidth]{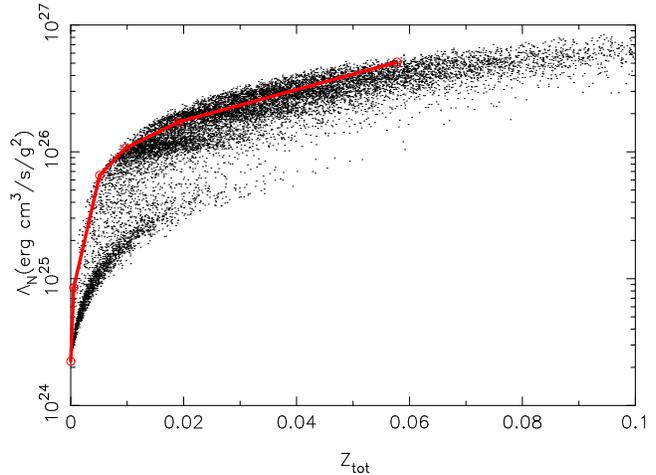}
 \caption{Cooling function dependence on the total metallicity $Z$ for a given ($T=10^{5.5}$ K) temperature for distribution of abundances present on a cosmological self-consistent simulation with metal enrichment. The solid line represents the cooling according to SD93.}\label{fig:cool1}
\end{figure}

In order to test the above approach, we have considered a sample of gas particles with different metal mixtures. Instead of assigning a random composition to each particle, we have randomly selected $\sim 10^5$ particles extracted from cosmological $\Lambda$CDM simulations with the chemical evolution model described in $\S$ \ref{sec:CEM}.

A full computation of the cooling function has been carried out for each particle by using MAPPINGS and its individual chemical composition. The dots in Fig. \ref{fig:cool1} represent the individual values of the cooling function when the same temperature ($T=10^{5.5}$) is assigned to all the gas particles in the sample. The solid line represents instead the cooling rate obtained from the SD93 method. It can be seen from this figure that the latter approach gives a reasonable approximation of the cooling rate at different $Z_{\rm{tot}}$ values. However, the figure also shows an important dispersion on the cooling rate of gas particles with the same total metallicity but different metal mixtures. Such a dispersion could lead to errors in the estimate of $\Lambda_T(\mathbf{Z})$ of almost one order of magnitude for sub-solar metallicities.

Different approaches can be envisaged to improve the modelling of the
cooling function in cosmological simulations. For example by
characterising the composition dependence of the cooling function
with more than one parameter. This could be achieved by considering,
in addition to the total metallicity, the alpha-element enhancement
or any other parameter providing a more detailed description of the
metal content. Another approach that needs to be explored consists of
maintaining the Eq. (\ref{CoolingRedux}), with just one
metallicity parameter, but using a different choice of $\zeta$ to
describe the effect on the cooling function of different chemical
compositions.

In order to analyse this latter possibility, we have employed a
Dimension Reduction Regression (DDR) technique \citep{2002JSS.....7....0W}.
In such a procedure, one tries to reduce the multidimensional
dependence of a function $\Lambda(\mathbf{Z})$ (e.g., the cooling
function at a given temperature) to a small number $d$ of parameters
expressed as linear combinations
$\zeta_1=\mathbf{c}_1\cdot\mathbf{Z},\, ...,\,
\zeta_d=\mathbf{c}_d\cdot\mathbf{Z}$, where $\mathbf{c}_i$ are
vectors, of the same dimension as $\mathbf{Z}$, containing the
coefficients of each linear combination $\zeta_i$. If $d$ is very
small, one or two, then the regression problem can be summarised
using simple graphics (a single $\Lambda(\zeta_1)$ plot for $d=1$,
or a 3D plot for $d=2$) containing all the regression information.

Several methods for estimating $d$ and the relevant coefficients
$\mathbf{c}_1,..., \mathbf{c}_d$ have been suggested in the
literature. In this paper we have used a sliced inverse regression
method \citep{1991JASA...86..316L}, where the range of $\Lambda_T$ values is
divided into several intervals, or slices, and then a weighted
principal component analysis is performed. This gives higher
importance to the slices with higher covariance. The eigenvectors
of the covariance matrix, ordered by their corresponding eigenvalues,
are the preferred directions (i.e. the $d$ $\mathbf{c}_i$ vectors).

When the above algorithm is applied to the cooling function, one finds that just one parameter
\begin{equation}\label{zetaT}
\zeta(T)=\mathbf{c}(T)\cdot\mathbf{Z}\;
\end{equation}
is enough to accurately fit $\Lambda_N$ at a given temperature $T$. 
For the sake of clarity, Table 1 gives the resulting $\mathbf{c}(T)$ coefficients for some temperature values and for the same metals as those considered in the yields of \citet{2005A&A...432..861G}. The metallicity parameter in Table 1 has been normalised according to $\zeta_N(T)=[\zeta(T)-\zeta_0(T)]\zeta_1(T)$ so that $\zeta_N(T)$ lies in the interval $[0,1]$. The corresponding $\zeta_0(T)$ and $\zeta_1(T)$ normalisation parameters are also given in Table 1. An extended version of this Table is available online. 

\begin{figure}
 \centering
 \includegraphics[width=\columnwidth]{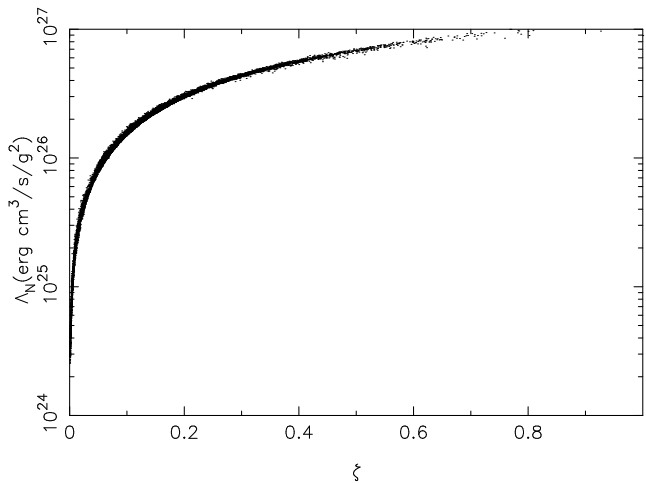}
 \caption{Cooling function dependence on the $\zeta_T$ parameter for a given ($T=10^{5.5}$ K) temperature.}\label{fig:cool2}
\end{figure}

To check the above result, we have repeated the same test as in
Fig. \ref{fig:cool1}. For each gas particle in the
sample we have considered its detailed chemical content to
determine its corresponding $\zeta$ parameter (Eq.
\ref{zetaT}), as well as to carry out a full MAPPINGS III computation of its
individual cooling rate at $T=10^{5.5}$ K. The dots in Fig.
\ref{fig:cool2} show the $(\Lambda_j,\zeta_j)$ values obtained for
each particle $j$. It can be seen that such a plot now presents
almost no dispersion and, therefore, the $\Lambda_N(T,\zeta)$ values
can be easily and accurately tabulated (see Table 2).

\begin{figure}
 \centering
 \includegraphics[width=\columnwidth]{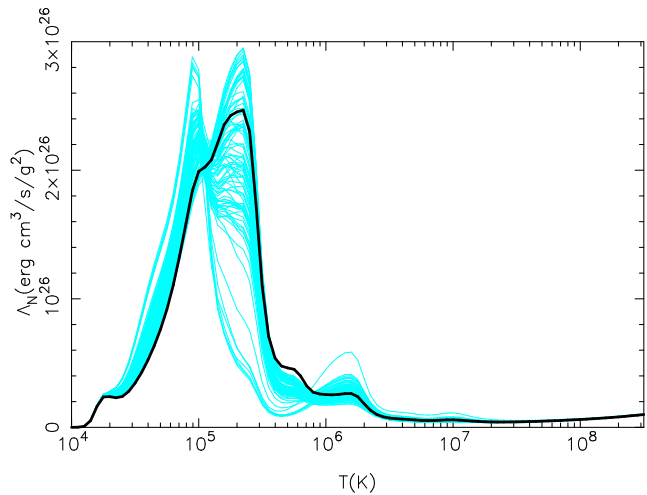}
 \caption{Cooling function for a sample of particles with solar metallicity
($Z=0.02\pm0.0002$) but different metal mixtures as found in a
cosmological simulation (thin lines). The thick black line represents the cooling
function for solar abundances \citep{1989GeCoA..53..197A}.} \label{fig:coolsol}
\end{figure}

Summarising, given a gas particle with known temperature $T$ and
chemical content $\mathbf{Z}$, its appropriate metallicity parameter
$\zeta(T,\mathbf{Z})$ can be computed by using Eq.
(\ref{zetaT}) and the coefficients of Table 1. An accurate estimate
of the cooling rate can be then obtained through interpolation from
Table 2. The resulting algorithm needs an almost negligible amount
of computing time but implies a remarkable improvement on the
cooling modelling, as compared to other approaches based on
$Z_{\rm{tot}}$. Indeed, we have applied this algorithm to a sample
of simulated gas particles with $Z_{\rm{tot}}=Z_{\sun}$ but
different chemical contents. Fig. \ref{fig:coolsol} shows their
corresponding cooling functions (thin lines), as well as the cooling
function (thick line) obtained for solar abundances (i.e., in
the case considered by SD93).

\begin{figure}
 \centering
 \includegraphics[width=\columnwidth]{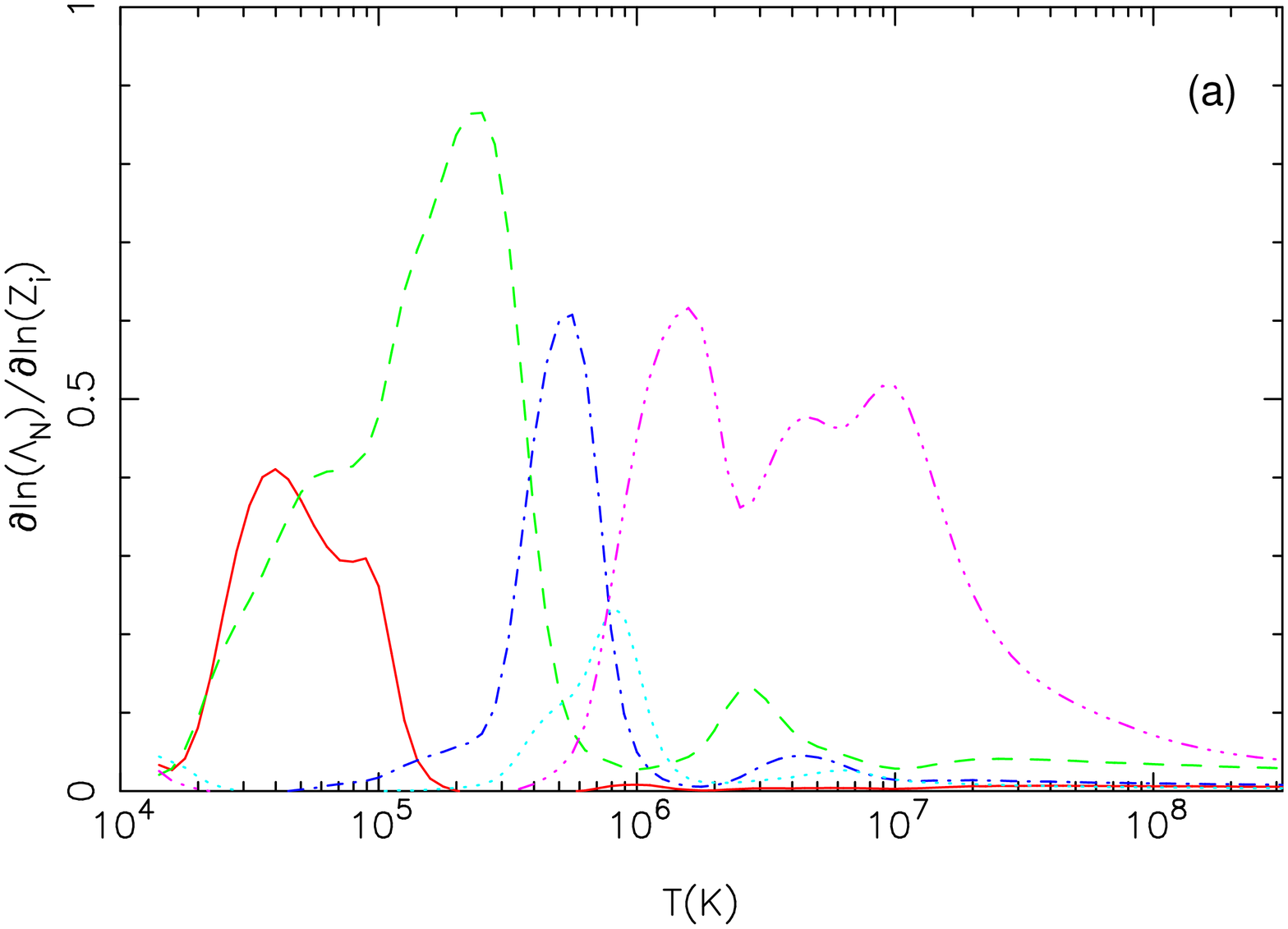}
 \includegraphics[width=\columnwidth]{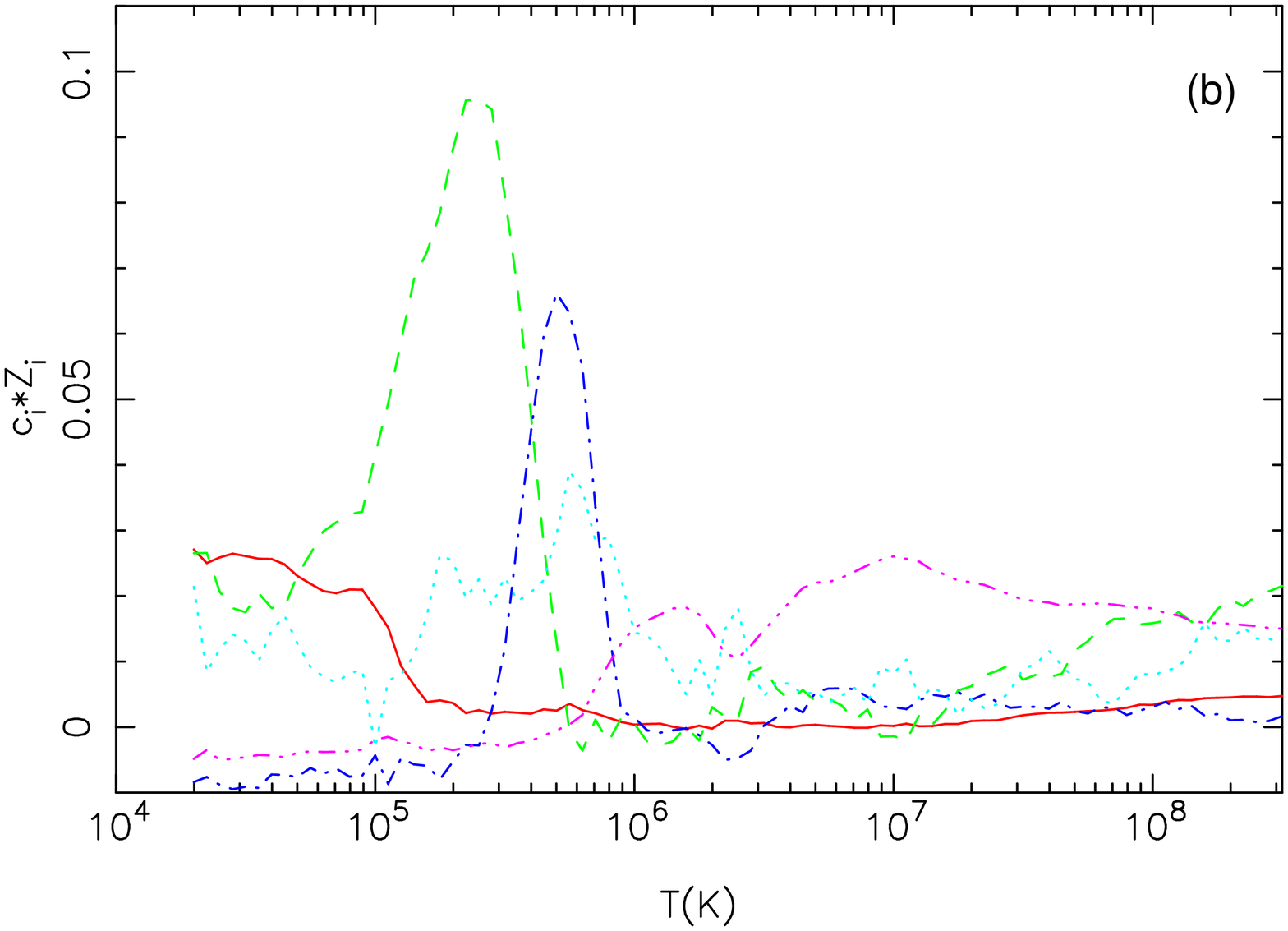}
 \caption{\textbf{(a)}: portion of the total cooling function contributed by
C (solid line), O (dashed line), Ne (dashed-dotted line), Mg (dotted
line) and Fe (dashed-triple dotted line) for a solar mixture of
elements ($\mathbf{Z} = \mathbf{Z}_{\sun}$). \textbf{(b)}:
corresponding values of the $c_i Z_i$ terms of
$\zeta_T$.}\label{fig:cpartial}
\end{figure}

The physical meaning of the $\mathbf{c}(T)$ values obtained in the
above DDR procedure can be understood as coefficients
giving, at a given temperature $T$, the weight of each element on
the total cooling function. For example, Fig. \ref{fig:cpartial}-a
shows the logarithmic partial derivative of
$\Lambda_N$ with respect to the abundance of each element, i.e., a
quantity roughly giving the weight of each element on the total
cooling function. Fig. \ref{fig:cpartial}-b shows instead
the contribution $\zeta_i=c_iZ_i$ of each element $i$ to the
metallicity parameter $\zeta(T)$. In both panels, solar proportions
have been assumed. It can be seen from this figure that both
quantities are correlated, with the most important coolants
contributing more to the metallicity parameter $\zeta(T)$.

It is important to note that, in SPH simulations, specific
internal energy $u$ is tracked instead of temperature $T$. Both are
related by
\begin{equation}
\label{eq:u_T}
u_T(\mathbf{Z})=\frac{3}{2\bar{\mu}m_p}k_BT\;,
\end{equation}
where $k_B$ is Boltzmann's constant, $m_p$ is the proton mass and $\bar{\mu}$ is the mean molecular mass. For simplicity, some SPH codes fix $\bar{\mu}$ to a constant value regardless of metallicity. In some cases \citep[e.g.,][]{2003MNRAS.340..908K} such a constant value is chosen to represent a fully ionised gas ($\bar{\mu}=0.6$), while in other cases \citep[e.g.,][]{1999A&A...348..371B} the adopted value represents a cool gas with primordial ($\bar{\mu}=1.2$) or solar ($\bar{\mu}=1.3$) abundances. To be consistent with our aim of developing a model that takes into account the full dependence on the chemical composition, we have preferred to consider the mean molecular mass as a function of $T$ and $\mathbf{Z}$. Consequently, we have applied a DDR procedure similar to that of Eq. (\ref{CoolingRedux}) to write $u(T,\mathbf{Z})$ as a linear combination of $\mathbf{Z}$. The corresponding DDR coefficients are also available online.

\section{Tests and results}
\label{sec:4}

\subsection{Synthetic tests}\label{sec:synthetic}

In order to test the results of our model, as well as the resolution effects, we have carried out a series of tests based on classical evolution models for chemical evolution. In all these tests, particles are just discrete mass elements submitted to certain constraints. Therefore the models presented in this section are useful to test the different aspects of our implementation of the metal enrichment, uncoupled from any dynamical effect.

\subsubsection{A single stellar population} \label{sec:SSP}

Following LPC02, a first basic test consists of analysing the chemical evolution in the simple case of a single burst of star formation. Such a test is designed to check the validity of our statistical approach by comparing its results to the expected chemical production of an SSP.

The single-burst model begins with $N=5000$ newly born star particles of total mass $10^{11}$ M$_\odot$. All particles have primordial composition at $t=0$ and are left to evolve afterwards according to the statistical prescriptions for gas restitution and metal enrichment described in $\S$ \ref{sec:CEM}. Consequently, as time progresses, some stellar particles will turn into gas. At each time step, the metal content of the gas component is enriched through the yields of the remaining stellar particles according to either Eq. (\ref{eq:newabund}) or (\ref{eq:spread}). In this test, metals are instantaneously mixed through the gas component and no further SF episodes take place.

\begin{figure*}
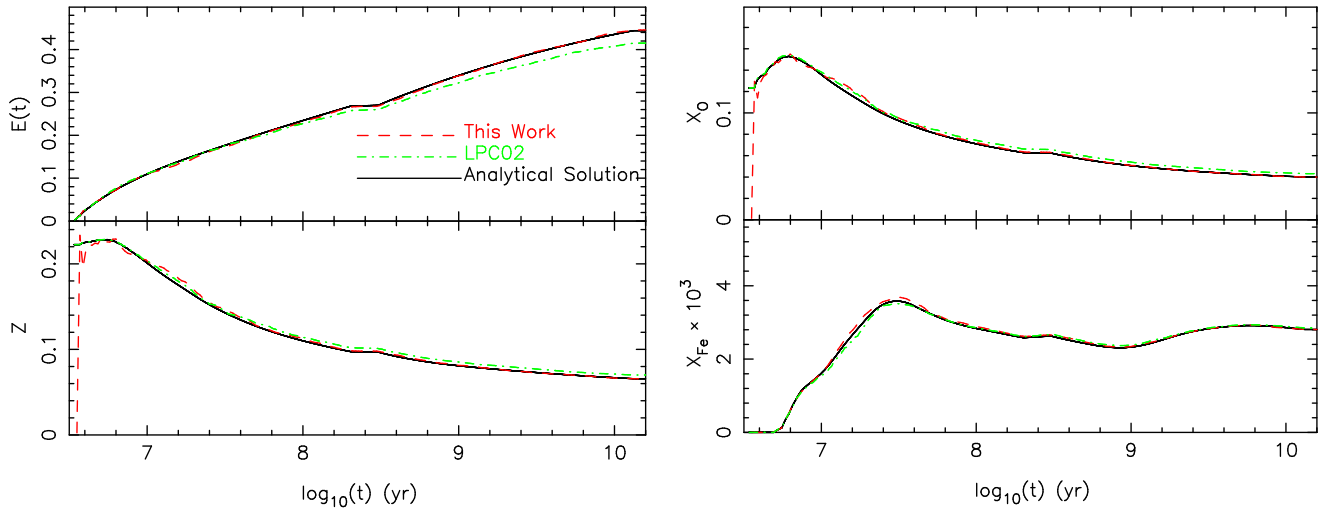

\begin{tabular}{cc}
 \includegraphics[angle=-90,width=0.48\textwidth]{fig6a} &
 \includegraphics[angle=-90,width=0.48\textwidth]{fig6b} \\
\end{tabular}
 \caption{Results of the SSP test: time evolution of the total gas ejection $E(t)$, total metallicity of the ejected gas, $Z$, oxygen and iron fraction in the ejected gas}\label{fig:testnpart}
\end{figure*}

Fig. \ref{fig:testnpart} displays, as a function of time, the gas mass fraction and the chemical composition of the gas. The solid lines correspond to the exact analytical predictions for an SSP (i.e., those obtained in the continuous limit by directly integrating Eqs. \ref{ET} and \ref{eq:spread} for a unique SSP). The dashed and dotted-dashed lines give instead the results obtained from a statistical model of chemical evolution based on Eq. (\ref{eq:spread}) and Eq. (\ref{eq:newabund}), respectively. It can be seen from this figure that both statistical methods closely reproduce the expected trends for an SSP. Therefore, our scheme of progressive metal ejection (Eq. \ref{eq:spread}) gives, within a stochastic model, a fair representation of the overall production of an SSP.

\subsubsection{Closed box model of an elliptical galaxy}\label{sec:closedbox}

Another useful test for chemical evolution models is the closed box model \citep{1997nceg.book.....P} with one zone. We have implemented such a test by initially considering $N$ gas particles that, during the simulation, randomly turn into stars according to a pre-defined probability. More specifically, in order to mimic an elliptical galaxy, we have considered an exponentially decaying star formation rate:
\begin{equation}
 \psi(t)=\kappa e^{-\tau t} \,,
 \label{eq:expsfr}
\end{equation}
so that, assuming no gas feedback, the rate of change of the gas mass fraction, $g(t)$, is given by
\begin{equation}
dg(t)/dt = - \psi(t)\,.
\end{equation}

The above equation implies that the probability $p$ that a gas particle forms stars in a time $\Delta t$ is
\begin{equation}
 p=\frac{\left(e^{-\tau\Delta t} -1\right)
 \kappa}{\kappa+e^{\tau t} (\tau-\kappa)}\,,
 \label{eq:toysfr2}
\end{equation}
where $\kappa$ can be written in terms of the final fraction of gas, $g_1$:
\begin{equation}
 \kappa=\frac{\tau e^\tau (g_1 -1)}{1 - e^\tau}\,.
 \label{eq:toysfr2kappa}
\end{equation}

As in the previous test, all particles are left to evolve according to the statistical model for gas restitution and metal enrichment described in $\S$ \ref{sec:CEM}, except that here we use Eq. (\ref{eq:toysfr2}) to compute $p$ at each time step for all gas particles and draw random numbers to decide which particles actually form stars. We again consider that metals are instantaneously mixed through the gas component. 

\begin{figure*}
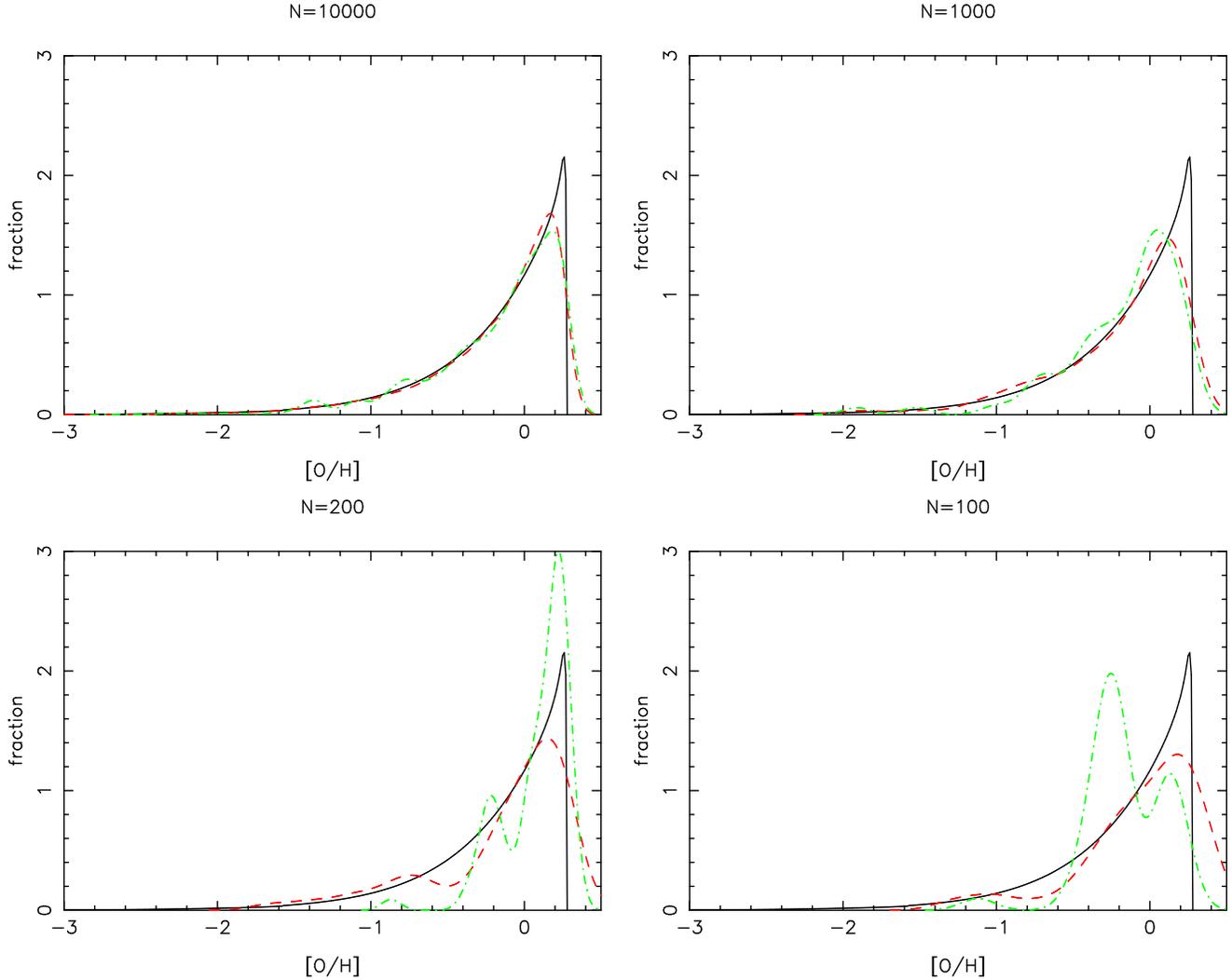

\begin{center}
\begin{tabular}{cc}
 \includegraphics[angle=-90,width=0.48\textwidth]{fig7a} &
 \includegraphics[angle=-90,width=0.48\textwidth]{fig7b} \\
 \includegraphics[angle=-90,width=0.48\textwidth]{fig7c} &
 \includegraphics[angle=-90,width=0.48\textwidth]{fig7d} \\
\end{tabular}

\caption{\label{fig:testclosed} Metallicity Distribution Function (MDF) for the stellar particles in the closed box test. The different panels correspond to a decreasing number of particles ($N=$10000, 1000, 200 and 100) and, therefore, to a decreasing mass resolution. In order to make easier the comparison of the different curves, the particle-based MDFs are displayed as kernel-smoothed distributions with the bandwidth, $h=1.06\sigma N^{-1/5}$, where $\sigma$ is the standard deviation of the sample, as suggested by \citet{1992mde..book.....S}. The solid line corresponds to the analytical solution, the dashed-line to the numerical results obtained from Eq. (\ref{eq:spread}), and the dotted-dashed line to those found from Eq. (\ref{eq:newabund}).}
\end{center}
\end{figure*}

The above closed box model then gives a simple representation of objects with a more complex sequence of star formation bursts and, therefore, containing a mixture of different SSPs. The main advantage of this model is that it has a known theoretical prediction for the resulting metal distribution function (MDF) of the gas component (see Appendix \ref{appendixa}). The comparison with such a theoretical MDF then constitutes a strong test, uncoupled from hydrodynamic and gravitational effects, of the stability of particle-based numerical methods against a degraded resolution. 

We have run this test for four different particle numbers ($N=10000$, 1000, 200 and 100). In all runs we considered $\tau = 1/5$ Gyr$^{-1}$, a typical value for massive elliptical galaxies, and a constant timestep of $\Delta t = 1.38\; 10^6$ yr (i.e., 10000 integration steps over a Hubble time). In order to have enough resolution for the gas component, we have used a high value ($g_1 = 0.4$) for the final gas fraction. In a series of runs we have used Eq. (\ref{eq:newabund}) to incorporate the stellar production of metals into the gas component, whereas in another series of runs we have used the new procedure proposed in this paper (Eq. \ref{eq:spread}) to account for the metal feedback. Fig. \ref{fig:testclosed} shows the results of this test, where the solid line corresponds to the theoretical MDF, whereas the particle-based predictions are shown as dashed (for Eq. \ref{eq:spread}) and dotted-dashed (for Eq.\ref{eq:newabund}) lines. As can be seen from Fig. \ref{fig:testclosed}, for a high number of particles ($N=10000$) both procedures for metal feedback lead to MDFs in good agreement with the theoretical solution. Obviously, the kernel procedure used to draw the particle-based results leads to MDFs where the sharp peak and cutoff of the theoretical solution at [O/H] $\sim$ 0.23 are smoothed. In tests with $N=1000$ particles, both metal feedback procedures still give results in reasonable agreement with the theoretical solution, although the statistical noise is slightly less visible in the predictions obtained from Eq. (\ref{eq:spread}) than in those found from Eq. (\ref{eq:newabund}). At low particle numbers ($N=200$ and 100), the results from Eq. (\ref{eq:spread}) degrade sensibly better and keep the shape of the MDF almost unaltered, while those from Eq. (\ref{eq:newabund}) become noise-dominated due to the lack of enough sampling data.

\subsubsection{Multi-zone model of a Spiral Galaxy}\label{sec:multi-zone}

In order to test our implementation of the Q$_{ij}$ formalism, we have also performed a test based on the \citet{2005MNRAS.358..521M} multi-zone evolution model for a disk galaxy. Such a model also considered the Q$_{ij}$ formalism and used the same stellar libraries as in our code. To ensure that our results are directly comparable to those of \citet{2005MNRAS.358..521M} and \citet{2005A&A...432..861G}, we have adopted the same IMF \citep{1990A&A...231..391F,1992ApJ...387..138F} for this test as that used in such studies.

The \citet{2005MNRAS.358..521M} model considers a galaxy with a spherical halo of radius $R_H$ and a concentric cylinder constituting the disk. The spherical halo is divided into concentric cylindrical regions 1 kpc wide with a height determined by the corresponding galactocentric distance on the disk and the total radius of the sphere. The corresponding regions on the disk are also concentric cylindrical shells 1 kpc wide, but with a constant height $h_D=0.2$ kpc.

In the initial conditions, the halo contains the total mass of the galaxy in gas phase. The halo mass has a radial distribution consistent with the rotation curve of \citet{1996MNRAS.281...27P}. In units of $10^9$ M$_\odot$, such a distribution is given by $M(R)=2.32\times 10^5 R V(R)^2$, with
\begin{equation}
V(R)=V_{opt}\left[ 0.72\frac{1.97x^{1.22}}{(x^2+0.61)^{1.43}}+1.07\frac{x^2}{x^2+2.25}\right]^{1/2}\,.
\end{equation}

Here, $x=R/R_{opt}$ and $R_{opt}=R_H/2.5$. We have considered $V_{opt}=200$ km/s, $R_{opt}=13$ kpc and $L=10^{10.4}L_\odot$, which correspond to galaxies similar to the Milky Way. 

For each region $i$ the matter in the halo can be either in the form of stars of diffuse gas, with total mass $M_{sh}$ and $M_{gh}$, respectively. In the corresponding disk region, the model allows for the following phases: diffuse gas ($M_{gd}$), clouds ($M_{cd}$), and stars ($M_{sd}$). In this latter phase, the adopted IMF implies a certain mass $M_{s2}$ of stars with $M>8$ M$_\odot$\footnote{Note that the mass of massive stars in an SSP, $M_{s2}$, declines with the SSP age and eventually drops to zero.}. The mass of the different phases can change through the following conversion processes: i) Diffuse gas infall from the halo to the disk, with a rate $\dot{m}_{INF}$, ii) In the halo, star formation from diffuse gas ($\dot{m}_{SFH}$), iii) In the disk, star formation from clouds ($\dot{m}_{SFD}$), either from cloud-cloud collisions and from massive star-cloud interactions, and iv) Cloud formation from diffuse gas in the disk ($\dot{m}_{CFD}$). The complete set of equations for each zone is therefore \citep[see][for details]{1996ApJ...466..668M}:
\begin{eqnarray}
dM_{gh}/dt&=&\dot{W}_H-\dot{m}_{INF}-\dot{m}_{SFH}\nonumber\\
dM_{sh}/dt&=&\dot{m}_{SFH}-\dot{W}_H\nonumber\\
dM_{sd}/dt&=&\dot{m}_{SFD}-\dot{W}_D\label{setMM}\\
dM_{cd}/dt&=&\dot{m}_{CFD}-\dot{m}_{SFD}\nonumber\\
dM_{gd}/dt&=&\dot{m}_{INF}-\dot{m}_{CFD}+\dot{W}_D\nonumber
\end{eqnarray}
where $W_H$ and $W_D$ represent the gas return rate for the halo and disk, respectively, and
\begin{eqnarray}
\dot{m}_{INF}&=&M_{gh}/\tau\nonumber\\
\dot{m}_{SFH}&=&K M_{gh}^{1.5}\nonumber\\
\dot{m}_{CFD}&=&\mu M_{gd}^{1.5}\label{MMprocesses}\\
\dot{m}_{SFD}&=&H M_{cd}^2+\epsilon_a M_{cd}M_{s2}\nonumber
\end{eqnarray}
with $\tau=\tau_c \exp[(R-R_c)/\lambda_D]$, $K = \epsilon_K (G/V_H)^{1/2}$, $\mu = \epsilon_\mu (G/V_D)^{1/2}$, and $H=\epsilon_H/V_D$. Here, $R_c=R_{opt}/2$, $\lambda_D=0.15R_{opt}$, $G$ is the universal gravitational constant, $V_H$ is the volume of the halo region and $V_D$ is the volume of the disk region. We have considered parameter values that correspond to a galaxy mimicking the Milky Way: $\tau_c=4$ Gyr, $\epsilon_K=5.3\,10^{-3}$, $\epsilon_\mu=0.80$, $\epsilon_H=0.28$ and $\epsilon_a=0.83$. 

The above model has been implemented by considering 16 regions in the disk, from $R = 2$ kpc to $R = 18$ kpc. We explicitely avoid the bulge since our model is not applicable there. Each region is sampled with
\begin{equation}
 N = 1000 \max\left[1,\,\frac{1}{1-\exp(-1/\tau)}\right]
\end{equation}
particles of individual mass $M(R)/N$. All particles are initially labelled as diffuse gas in the halo so that, except for $M_{gh}$, all phases have a vanishing total mass. At each timestep $\Delta t$, the rate of change for the mass of the different phases is computed using Eqs. (\ref{setMM})-(\ref{MMprocesses}). The corresponding change on the mass $M_i$ of each phase $i$ is then computed at the first level of approximation, $\Delta M_i=(dM_i/dt)\Delta t$, and then expressed as discrete changes in the numbers of particles. After re-labelling the particles, our routines for chemical evolution are called to compute the metal production and return of gas (terms $W_H$ and $W_D$ in Eqs. \ref{setMM}) in both the halo and the disk, except for the star formation, which is computed in this text from Eq. (\ref{MMprocesses}).

\begin{figure}
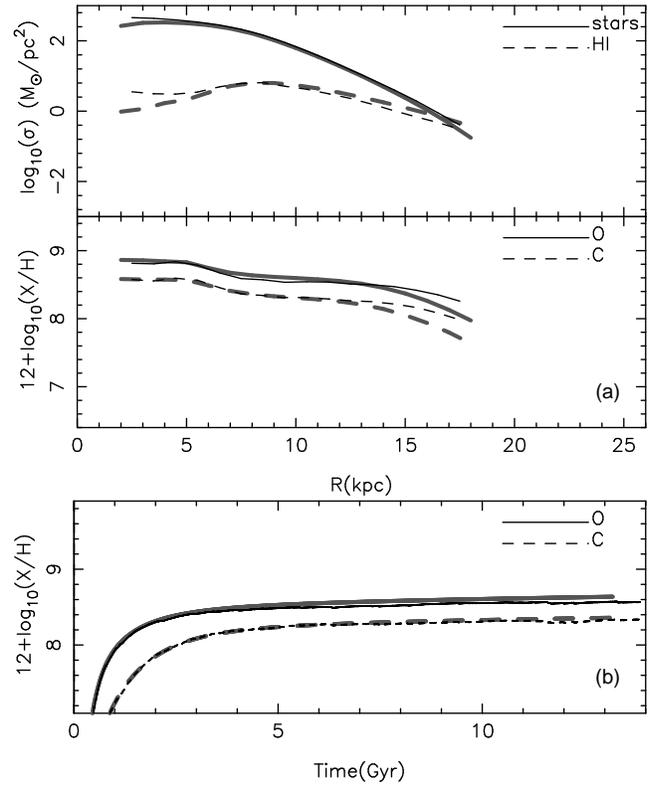

 \centering
 \includegraphics[angle=-90,width=\columnwidth]{fig8a}\\
 \includegraphics[angle=-90,width=\columnwidth]{fig8b}
 \caption{Comparison of our results (thin lines) and those found by \citet{2005MNRAS.358..521M} (thick lines) in their multi-zone model of a spiral galaxy. \textbf{(a)} Top panel: radial distributions of atomic gas, molecular clouds and stellar density at the end of the run. Bottom panel: radial distributions of oxygen and carbon. \textbf{(b)}: Time evolution of oxygen and carbon abundances in the gas for the solar cylinder.}\label{fig:testmm}
\end{figure}

Fig. \ref{fig:testmm} compares our results and those found by \citet{2005MNRAS.358..521M} at $t=13.72$ Gyr. We see from this figure that the radial distribution of the different elements and model components obtained from our code are in close agreement with those of \citet{2005MNRAS.358..521M}. The small deviation of the oxygen and carbon abundance at the outer galactic regions is probably due to the fact that our approach is based on masses that are sampled by a discrete number of particles. Both methods give also very similar results for the time evolution of the oxygen and carbon abundance in the gas for the solar cylinder.

\begin{figure}
 \centering
  \includegraphics[angle=-90,width=\columnwidth]{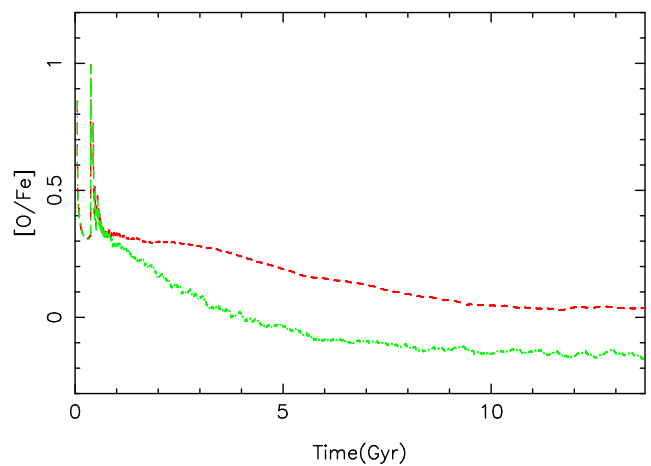}
\caption{Time evolution of the [O/Fe] ratio of stellar particles in the solar cylinder. The dashed line corresponds to the result obtained from the Q$_{ij}$ formalism. The result displayed as a dashed-dotted line has been obtained by computing the metal production of each stellar particle using its total metallicity and assuming solar proportions.}\label{fig:ofemmt}
\end{figure}

Fig. \ref{fig:ofemmt} shows the time evolution of the stellar [O/Fe] ratio obtained from the Q$_{ij}$ formalism (dashed line) as compared to that found when the Q$_{ij}$ formalism is not used (dashed-dotted line). This latter result has been computed, for each stellar particle with composition $X_j$, by using in Eq. (\ref{eq:ejection}) effective abundances $X'_j$ scaled, assuming solar proportions, according to its total metallicity. We see from this figure that the latter procedure underestimates the [O/Fe] ratio. This is due to the fact that forcing solar proportions for a stellar particle with [O/Fe]$>0$ implies the use in Eq. (\ref{eq:ejection}) of a larger effective abundance of iron ($X'_{\rm{Fe}}>X_{\rm{Fe}}$). Consequently, when computing the ejection of iron $e_{\rm{Fe}}(t)$, the dominant affected term in Eq. (\ref{eq:ejection}) is $q_{\rm{Fe,Fe}}X'_{\rm{Fe}}>q_{\rm{Fe,Fe}}X_{\rm{Fe}}$, and $e_{\rm{Fe}}(t)$ is overestimated. Although less significantly, similar arguments imply that the ejection of oxygen is underestimated. These two effects can also be noticed in Fig. \ref{fig:yields} and result in the underestimation of the [O/Fe] ratio observed in Fig. \ref{fig:ofemmt}.

\subsubsection{Diffusion in a homogeneus box} \label{sec:testdif}

The diffusion of metals in our model remains to be tested. In order to test our implementation of Eq. (\ref{eqn:diff}), described in $\S$ \ref{sec:diffusion}, we have devised a simple test. Starting with an isolated cubic box of $L=1.43$ Mpc filled with $32^3$ particles, distributed on a regular grid and representing a cloud of homogeneous gas with density $\rho=1.058\times10^{-35}$ g/cm$^3$, we pollute the central particle with a metallicity of $Z=0.1$, and let diffusion algorithm act for 13.72 Gyr. A diffusion constant of $D=4.63\times 10^{29}$ cm$^2$/s has been used, and no other processes have been considered, so that the particles are kept fixed to the grid.

This simple setup allows for an analytical solution of the diffusion equation with initial conditions
\begin{equation}
 X(\mathbf{r},t=0)=0.1 \delta(\mathbf{r})\;,
\label{eq:testdifini}
\end{equation}
namely
\begin{equation}
X(\mathbf{r},t) = \int X(\mathbf{r'},0) \frac{1}{(4\pi D t)^{3/2}}e^{-(\mathbf{r}-\mathbf{r'})^2/4 D t} \mathrm{d}\mathbf{r'}^3\;.
\end{equation}
While initial conditions of Eq. (\ref{eq:testdifini}) cannot be imposed in a discrete model, both should converge if $r$ or $t$ are big enough.

\begin{figure}
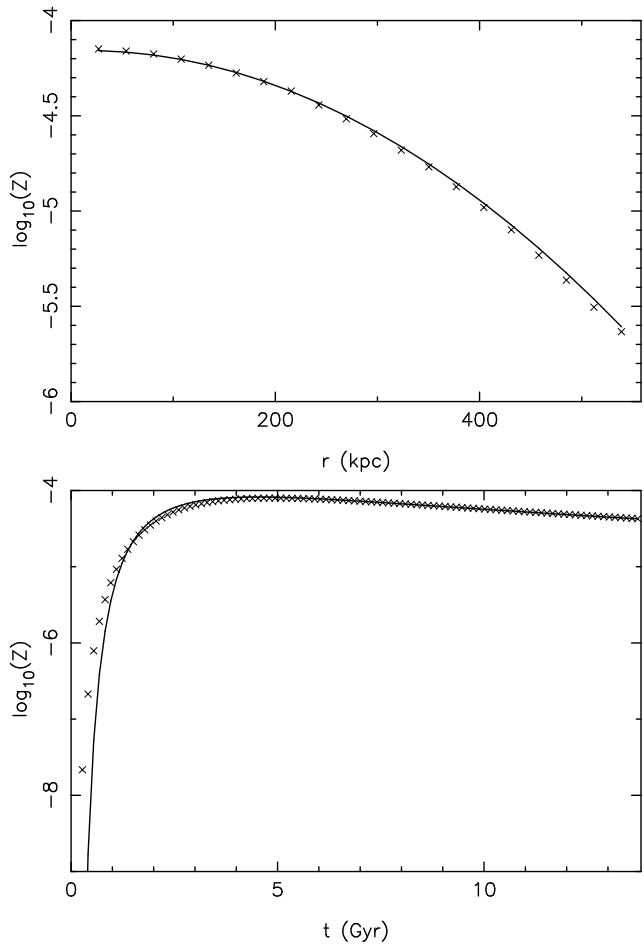

 \centering
  \includegraphics[angle=-90,width=\columnwidth]{fig10a}
  \includegraphics[angle=-90,width=\columnwidth]{fig10b}
\caption{Upper panel: final spherical distribution of metallicity for the diffusion test (crosses) and analytical solution (solid line) for $t=13.72$ Gyr. Lower panel: time evolution of the metallicity at $r=200$ kpc.}\label{fig:testdif}
\end{figure}

In fact, the upper panel of Fig. \ref{fig:testdif} shows the comparison between the final metallicity profile of the numerical test and its corresponding analytical solution. We find a good agreement for such a high $t$. Lower panel shows the metallicity time evolution in a bin centered around $r=200$ kpc. We find the numerical solution to have a shallower growth than the analytical one, due to the fact that the numerical initial conditions are not an exact delta function. At later times, however, both solutions converge.

\subsection{Cosmological simulations} \label{sec:cosmo}

We have finally performed different simulations of galaxy formation and chemical enrichment within a cosmological context. All simulations started at redshift $z= 20$ from initial conditions that are a Montecarlo realisation of the field of primordial fluctuations to a concordance cosmological model (a flat $\Lambda$CDM model, with $h=0.7$, $\Omega_{\rm \Lambda} = 0.7$, $\Omega_{\rm m} = 0.3$ and $\Omega_{\rm b} = 0.04$). The $\sigma_8$ value has been taken slightly high ($\sigma_8 = 1$) in order to mimick an active region of the universe \citep{1990ApJ...365...13E}. The evolution of these fluctuations was numerically followed up to $z =0$ by means of the parallel version of the DEVA code \citep[see][for a detailed description of DEVA]{2003ApJ...597..878S} where we have coupled our model for chemical evolution and cooling. In the test performed in this section we have used a star-formation threshold of $\rho_{th}=3\times10^{-25}$ g/cm$^3$, an efficiency of $c_\ast = 0.3$, a gravitational softening of $\epsilon = 1.5$ kpc/$h$ and $N=2\times64^3$ particles in a box of $L = 7$ Mpc$/h$, implying a mass resolution of $m_b = 2.06\times 10^7$ M$_\odot$ for baryonic particles and $m_{dm} = 1.34\times 10^8$ M$_\odot$ for dark matter particles.

Individual galaxy-like objects of different morphologies naturally appear in these simulations as a consequence of the cosmic mass assembly and the physical processes taken into account. Differently from the semi-analytical models of $\S$ \ref{sec:synthetic}, no assumptions are made about the spatial variations of the infall rate of gas and its relative importance as compared to the star formation rate.

These simulations then constitute an appropriate test to analyse: i) whether the resulting galaxy-like objects have realistic chemical properties; ii) the possible effects of including the cooling function presented in $\S$ \ref{sec:cooling}, that takes into account the full dependence on the detailed chemical composition of gas particles; and iii) the importance of using the Q$_{ij}$ formalism in cosmological simulations. In order to address these issues we have carried out different runs. In a first run (referred as the $\zeta$-cooling run) we used the composition-dependent cooling of $\S$ \ref{sec:cooling}, whereas in a second run (referred as the $Z_{\rm{tot}}$-cooling run) we used the total metallicity-dependent cooling of SD93. In both cases, we used the Q$_{ij}$ formalism of $\S$ \ref{sec:CEM}. In a third run (referred as the no-Q$_{ij}$ run), the metal production of each stellar particle was instead computed by using its total metallicity and assuming solar proportions. 

Elliptical-like objects (ELOs) constitute a family of objects with simple scaling relations \citep[see][for a study of the origin of the Fundamental Plane of ELOs obtained in DEVA simulations]{2005ApJ...632L..57O}. In this work, we focus on the study of ELOs because given their simplicity, less resolution is needed to properly simulate them. ELOs were identified as those objects having a prominent stellar spheroidal component with hardly disks at all. Here we will focus on the analysis of the chemical properties of the most massive ELOs at $z=0$, sampled with at least 1000 baryonic particles. This selection criterion produces eight massive ELOs, with a stellar mass range of $\sim 3\times10^{10} -1.5 \times 10^{11}$ M$_\odot$.

\begin{figure}
\begin{center}
 \includegraphics[angle=-90,width=\columnwidth]{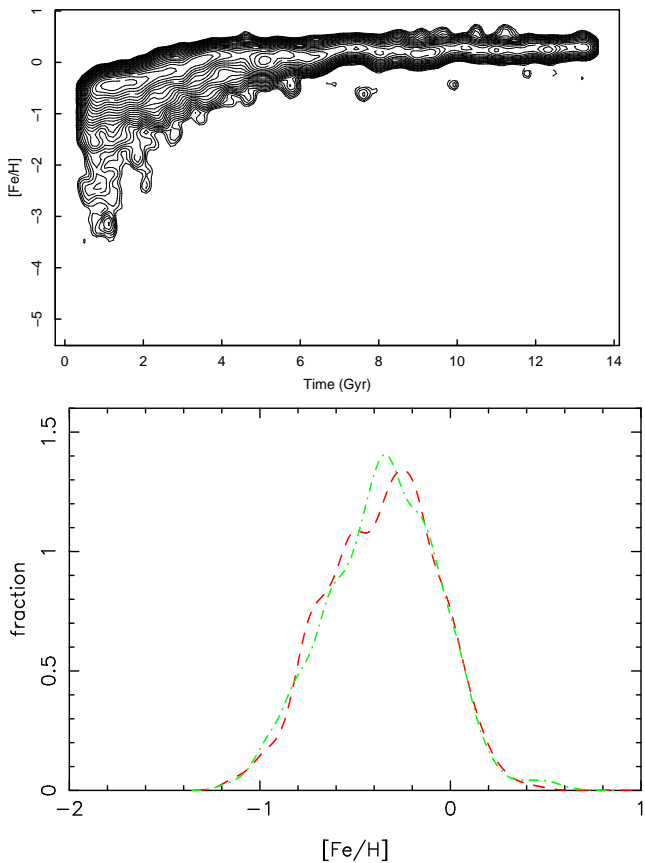}\\
 \includegraphics[angle=-90,width=\columnwidth]{fig11b}
\caption{\label{fig:testelip1} Chemical distribution for the most massive ELO in our simulation. The upper panel shows the age-metallicity relation as a contour plot. Only the $\zeta$-cooling run is displayed since there are virtually no differences between the relations arising from both cooling methods. The lower panel shows the stellar metallicity distribution function (number fraction of stars with a given metallicity) obtained in the $\zeta$-cooling (dashed line) and $Z_{\rm{tot}}$-cooling runs (dashed-dotted line) for RGB stars in the halo (see \citealt{2007AJ....134...43H} for an observational counterpart).}
\end{center}
\end{figure}

In both the $\zeta$-cooling and $Z_{\rm{tot}}$-cooling runs, we found that ELOs have metallicities with individual mean values that are consistent with those observed for elliptical galaxies with similar masses \citep{2005ApJ...621..673T}. For example, Fig. \ref{fig:testelip1} shows the age-metallicity relation and metallicity distribution function (MDF) for the most massive ELO ($M = 1.5\times 10^{11}$ M$_\odot$) identified in such simulations. We have found that the age-metallicity relation obtained for this object in the $\zeta$-cooling (shown in the upper panel of Fig. \ref{fig:testelip1}) is almost indistinguishable from that found in the $Z_{\rm{tot}}$-cooling run. Such a relation shows a very fast enrichment on the first 3 Gyr, when most of the star formation happens, followed by a very slow enrichment at recent times. The large scatter at $t<3$ Gyr is probably due to the fact that stars were formed on separate smaller objects that merged later on to form the final object. The MDFs found for this object in both simulations (lower panel of Fig. \ref{fig:testelip1}) are also very similar and with a shape that closely resembles that observed for some giant ellipticals \citep[e.g.,][]{2007AJ....134...43H}. In addition, the central oxygen abundance (not shown in Fig. \ref{fig:testelip1}) of this ELO has a value ([O/H]$\sim +0.2$) that agrees with that expected from the mass-metallicity relation given by the eq. (3) of \citet{2005ApJ...621..673T}

\begin{figure}
\begin{center}
\includegraphics[angle=-90,width=\columnwidth]{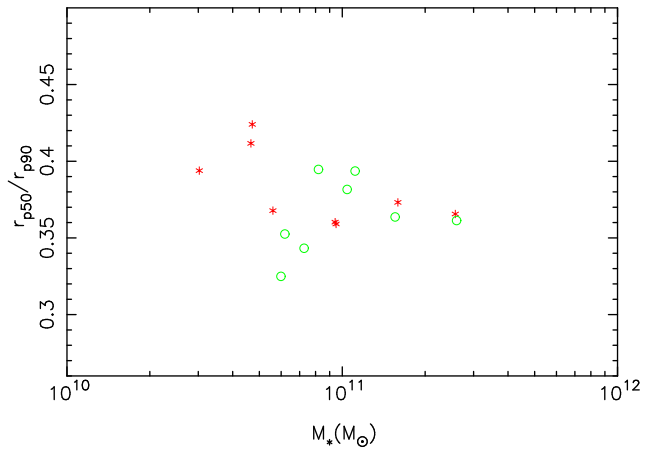}
\caption{\label{fig:testelip2} Inverse concentration index $C^{-1}=r_{p50}/r_{p90}$ as a function of total stellar mass of the ELOs found in the $Z_{\rm{tot}}$-cooling (circles) and $\zeta$-cooling (stars) runs.}
\end{center}
\end{figure}

In less massive ELOs, some significant differences appear between the results obtained from the $\zeta$ and $Z_{\rm{tot}}$ cooling runs. Fig. \ref{fig:testelip2} shows the inverse concentration index $C^{-1}=r_{p50}/r_{p90}$ (i.e., the ratio of the half-mass Petrosian radius to the 90\% mass Petrosian radius). It can be seen from this figure that the $Z_{\rm{tot}}$-cooling method produces objects that are more concentrated and with a higher stellar content than those found in the $\zeta$-cooling run. Such differences are small for the most massive objects but become important as we consider less massive ELOs. They are probably due to the fact that the $Z_{\rm{tot}}$ method has some tendency to overestimate the cooling rate of particles not having solar abundance ratios (see Fig. \ref{fig:cool1}).

\begin{figure}
\begin{center}
\includegraphics[angle=-90,width=\columnwidth]{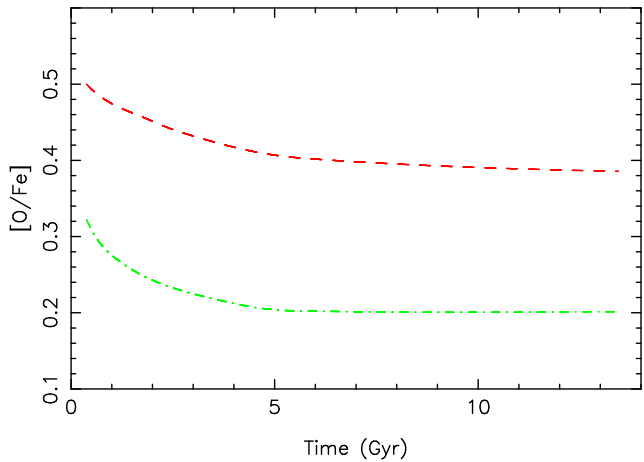}
\caption{\label{fig:qvsnoqelip} Time evolution of the stellar [O/Fe] ratio in the most massive ELO. The dashed line corresponds to the result obtained from the Q$_{ij}$ formalism, whereas the dotted-dashed line displays the result obtained in the no-Q$_{ij}$ run. In both cases, the $\zeta$-cooling method was used.}
\end{center}
\end{figure}

The effects of using the Q$_{ij}$ formalism in cosmological simulations are shown in Fig. \ref{fig:qvsnoqelip}. Such a figure displays, for the most massive ELO, the time evolution of the [O/Fe] ratio of stellar particles with radial distances $r<15$ kpc. The dotted-dashed line corresponds to the result obtained from the no-Q$_{ij}$ run, whereas the dashed line displays the result obtained when using the Q$_{ij}$ formalism and the same cooling method as in the no-Q$_{ij}$ run (i.e., the $\zeta$-cooling run discussed above). We see from this figure that, just like in the multi-zone test of $\S$ \ref{sec:multi-zone}, the assumption of solar proportions leads to a significant underestimation of the [O/Fe] ratio. 

\section{Summary and Conclusions}
\label{sec:5}

In this paper we have introduced an SPH model for chemical enrichment and radiative cooling in cosmological simulations of structure formation. Particular attention has been paid to including, by means of fast algorithms, the full dependence of both processes on the detailed chemical composition of star and gas particles. 

As compared with previous implementations in N-body simulations codes, the main features of our SPH-model for chemical evolution and cooling are:

i) Our model takes into account the delayed gas restitution from stars by means of a probabilistic approach that shares many aspects with that proposed by LPC02, except that the stellar yields are progressively distributed through the neighbouring gas (Eq. \ref{eq:spread}). The tests described in $\S$ \ref{sec:SSP} and \ref{sec:closedbox} show that our scheme, as well as that of LPC02 (based on Eq. \ref{eq:newabund}), provides a fair representation of the chemical production of an SSP and reproduces the expected trends for both one-zone and multi-zone models of galaxies. Nevertheless, a scheme based on our Eq. (\ref{eq:spread}) reduces the statistical noise and, at low particle numbers, degrades sensibly better than one based on Eq. (\ref{eq:newabund}), which becomes noise-dominated due to the lack of enough sampling data.

 ii) The full dependence of the metal production on the detailed chemical composition of stellar particles is taken into account by means of the Q$_{ij}$ formalism \citep{1973ApJ...186...51T}, that relates each nucleosynthetic product to its sources. Therefore, the assumption of solar proportions is relaxed in our model. Although the Q$_{ij}$ formalism has been previously considered in chemical evolution models \citep[e.g.,][]{1992ApJ...387..138F,2005A&A...432..861G}, this is the first time that it is used in N-body simulation codes. Releasing the assumption of solar proportions is important to accurately follow the evolution of abundance ratios like, e.g., the enhanced [$\alpha$/Fe] ratio observed in elliptical galaxies and in spiral galaxy bulges. This ratio, as well as its dependence on the central velocity dispersion, could provide us with important constraints on galaxy formation models. Indeed, the tests performed in this paper for both multi-zone models ($\S$ \ref{sec:multi-zone}) and cosmological simulations ($\S$ \ref{sec:cosmo}) suggest that the assumption of solar proportions leads to a significant underestimation of the [$\alpha$/Fe] ratio in simulated galaxy-like objects.

iii) In the same way, the full dependence of radiative cooling on the detailed chemical composition of gas particles has been implemented through a fast algorithm based on a metallicity parameter, $\zeta(T)$, that takes into account the weight of the different elements on the total cooling function. We have compared the results obtained when such a composition-dependent cooling is used (referred as $\zeta$-cooling runs) and those found from the total metallicity cooling tables of SD93 (referred as $Z_{\rm{tot}}$-cooling runs). To that end, we have carried out ($\S$ \ref{sec:cosmo}) different simulations of galaxy formation in the framework of a concordance cosmological model. For the most massive elliptical-like objects (ELOs), we have found that the differences between the $\zeta$ and $Z_{\rm{tot}}$-cooling runs were small. Such massive ELOs are characterised by strong and short bursty star formation events at early times, where most gas is exhausted and, therefore, where dynamics soon becomes dissipationless. However, for less massive ELOs, some important differences appear between the results obtained from the $\zeta$ and $Z_{\rm{tot}}$-cooling runs. Probably due to the fact that the $Z_{\rm{tot}}$ method has some tendency to overestimate the cooling rate of particles not having solar abundance ratios, such a method produces ELOs that are more concentrated and with a higher stellar content than those obtained in the $\zeta$-cooling run. Such differences become larger as we consider ELOs with smaller stellar masses. 

The above scheme for chemical enrichment and cooling has been implemented in the parallel-OpenMP version of DEVA \citep{2003ApJ...597..878S}, a Lagrangian code particularly designed to study galaxy assembly in a cosmological context. In this code, gravity is computed through an AP3M-like method, while hydrodynamics is computed through an SPH technique with algorithms and correction terms ensuring an accurate implementation of conservation laws (energy, entropy and angular momentum). The DEVA code has already been able to produce both elliptical \citep{2004ApJ...611L...5D, 2006ApJ...636L..77D, 2005ApJ...632L..57O, 2006MNRAS.373..503O, 2007MNRAS.376...39O} and spiral \citep{2002Ap&SS.281..309S} galaxy-like objects with realistic structural, kinematic and dynamical properties. Using this model for chemical evolution and cooling, we will analyse in forthcoming papers some important chemical properties like, e.g., the metallicity distribution functions (MDF) and abundance gradients within individual objects of different morphologies and environments, as well as the fundamental metallicity relations (mass-metallicity and mass-[$\alpha$/Fe] ratio) for large samples of ELOs obtained in cosmological simulations.

\section*{Acknowledgements}

We thank the anonymous referee of this paper for his/her very valuable comments and suggestions. MM thanks Ruiz-Lapuente and coworkers for providing her with a numerical table of their model for supernova rates. FJMS was supported by the Spanish MEC under FPU grant AP2003-0279. This project was partially supported by the MCyT (Spain) through grants AYA2006-15492-C03-01 and AYA2006-15492-C03-02 and also by the regional government of Madrid through the ASTROCAM Astrophysics network (S--0505/ESP--0237).

\bibliography{ms}

\begin{thebibliography}{}

\bibitem[\protect\citeauthoryear{{Anders} \& {Grevesse}}{{Anders} \&
  {Grevesse}}{1989}]{1989GeCoA..53..197A}
{Anders} E.,  {Grevesse} N.,  1989, \gca, 53, 197

\bibitem[\protect\citeauthoryear{{Bell}, {McIntosh}, {Katz} \&
  {Weinberg}}{{Bell} et~al.}{2003}]{2003ApJS..149..289B}
{Bell} E.~F.,  {McIntosh} D.~H.,  {Katz} N.,    {Weinberg} M.~D.,  2003, \apjs,
  149, 289

\bibitem[\protect\citeauthoryear{{Berczik}}{{Berczik}}{1999}]{1999A&A...348..3%
71B}
{Berczik} P.,  1999, \aap, 348, 371

\bibitem[\protect\citeauthoryear{{Berczik} \& {Petrov}}{{Berczik} \&
  {Petrov}}{2001}]{2001KFNT...17..213B}
{Berczik} P.~P.,  {Petrov} M.~I.,  2001, Kinematika i Fizika Nebesnykh Tel, 17,
  213

\bibitem[\protect\citeauthoryear{{Boissier} \& {Prantzos}}{{Boissier} \&
  {Prantzos}}{1999}]{1999MNRAS.307..857B}
{Boissier} S.,  {Prantzos} N.,  1999, \mnras, 307, 857

\bibitem[\protect\citeauthoryear{{Boissier} \& {Prantzos}}{{Boissier} \&
  {Prantzos}}{2000}]{2000MNRAS.312..398B}
{Boissier} S.,  {Prantzos} N.,  2000, MNRAS, 312, 398

\bibitem[\protect\citeauthoryear{{Burstein}, {Bertola}, {Buson}, {Faber} \&
  {Lauer}}{{Burstein} et~al.}{1988}]{1988ApJ...328..440B}
{Burstein} D.,  {Bertola} F.,  {Buson} L.~M.,  {Faber} S.~M.,    {Lauer} T.~R.,
   1988, \apj, 328, 440

\bibitem[\protect\citeauthoryear{{Burstein}, {Faber}, {Gaskell} \&
  {Krumm}}{{Burstein} et~al.}{1984}]{1984ApJ...287..586B}
{Burstein} D.,  {Faber} S.~M.,  {Gaskell} C.~M.,    {Krumm} N.,  1984, \apj,
  287, 586

\bibitem[\protect\citeauthoryear{{Carollo}, {Danziger} \& {Buson}}{{Carollo}
  et~al.}{1993}]{1993MNRAS.265..553C}
{Carollo} C.~M.,  {Danziger} I.~J.,    {Buson} L.,  1993, \mnras, 265, 553

\bibitem[\protect\citeauthoryear{{Carraro}, {Lia} \& {Chiosi}}{{Carraro}
  et~al.}{1998}]{1998MNRAS.297.1021C}
{Carraro} G.,  {Lia} C.,    {Chiosi} C.,  1998, \mnras, 297, 1021

\bibitem[\protect\citeauthoryear{{Chabrier}}{{Chabrier}}{2003a}]{2003PASP..115%
..763C}
{Chabrier} G.,  2003a, \pasp, 115, 763

\bibitem[\protect\citeauthoryear{{Chabrier}}{{Chabrier}}{2003b}]{2003ApJ...586%
L.133C}
{Chabrier} G.,  2003b, \apjl, 586, L133

\bibitem[\protect\citeauthoryear{{Chiappini}, {Matteucci} \&
  {Gratton}}{{Chiappini} et~al.}{1997}]{1997ApJ...477..765C}
{Chiappini} C.,  {Matteucci} F.,    {Gratton} R.,  1997, \apj, 477, 765

\bibitem[\protect\citeauthoryear{{Clayton}}{{Clayton}}{1987}]{1987ApJ...315..4%
51C}
{Clayton} D.~D.,  1987, ApJ, 315, 451

\bibitem[\protect\citeauthoryear{{Davidge} \& {Clark}}{{Davidge} \&
  {Clark}}{1994}]{1994AJ....107..946D}
{Davidge} T.~J.,  {Clark} C.~C.,  1994, \aj, 107, 946

\bibitem[\protect\citeauthoryear{{D\'{\i}az} \& {Tosi}}{{D\'{\i}az} \&
  {Tosi}}{1986}]{1986A&A...158...60D}
{D\'{\i}az} A.~I.,  {Tosi} M.,  1986, A\&A, 158, 60

\bibitem[\protect\citeauthoryear{{Dom{\'{\i}}nguez-Tenreiro}, {O{\~n}orbe},
  {S{\'a}iz}, {Artal} \& {Serna}}{{Dom{\'{\i}}nguez-Tenreiro}
  et~al.}{2006}]{2006ApJ...636L..77D}
{Dom{\'{\i}}nguez-Tenreiro} R.,  {O{\~n}orbe} J.,  {S{\'a}iz} A.,  {Artal} H.,
    {Serna} A.,  2006, \apjl, 636, L77

\bibitem[\protect\citeauthoryear{{Dom{\'{\i}}nguez-Tenreiro}, {S{\'a}iz} \&
  {Serna}}{{Dom{\'{\i}}nguez-Tenreiro} et~al.}{2004}]{2004ApJ...611L...5D}
{Dom{\'{\i}}nguez-Tenreiro} R.,  {S{\'a}iz} A.,    {Serna} A.,  2004, \apjl,
  611, L5

\bibitem[\protect\citeauthoryear{{Evrard}, {Silk} \& {Szalay}}{{Evrard}
  et~al.}{1990}]{1990ApJ...365...13E}
{Evrard} A.~E.,  {Silk} J.,    {Szalay} A.~S.,  1990, \apj, 365, 13

\bibitem[\protect\citeauthoryear{{Ferrini}, {Matteucci}, {Pardi} \&
  {Penco}}{{Ferrini} et~al.}{1992}]{1992ApJ...387..138F}
{Ferrini} F.,  {Matteucci} F.,  {Pardi} C.,    {Penco} U.,  1992, \apj, 387,
  138

\bibitem[\protect\citeauthoryear{{Ferrini}, {Moll\'{a}}, {Pardi} \&
  {D{\'{\i}}az}}{{Ferrini} et~al.}{1994}]{1994ApJ...427..745F}
{Ferrini} F.,  {Moll\'{a}} M.,  {Pardi} M.~C.,    {D{\'{\i}}az} A.~I.,  1994,
  ApJ, 427, 745

\bibitem[\protect\citeauthoryear{{Ferrini}, {Penco} \& {Palla}}{{Ferrini}
  et~al.}{1990}]{1990A&A...231..391F}
{Ferrini} F.,  {Penco} U.,    {Palla} F.,  1990, \aap, 231, 391

\bibitem[\protect\citeauthoryear{{Freeman} \& {Bland-Hawthorn}}{{Freeman} \&
  {Bland-Hawthorn}}{2002}]{2002ARA&A..40..487F}
{Freeman} K.,  {Bland-Hawthorn} J.,  2002, \araa, 40, 487

\bibitem[\protect\citeauthoryear{{Fuhrmann}}{{Fuhrmann}}{1999}]{1999Ap&SS.265.%
.265F}
{Fuhrmann} K.,  1999, \apss, 265, 265

\bibitem[\protect\citeauthoryear{{Galli}, {Palla}, {Ferrini} \&
  {Penco}}{{Galli} et~al.}{1995}]{1995ApJ...443..536G}
{Galli} D.,  {Palla} F.,  {Ferrini} F.,    {Penco} U.,  1995, \apj, 443, 536

\bibitem[\protect\citeauthoryear{{Gavil{\'a}n}, {Buell} \&
  {Moll{\'a}}}{{Gavil{\'a}n} et~al.}{2005}]{2005A&A...432..861G}
{Gavil{\'a}n} M.,  {Buell} J.~F.,    {Moll{\'a}} M.,  2005, \aap, 432, 861

\bibitem[\protect\citeauthoryear{{Gavil{\'a}n}, {Moll{\'a}} \&
  {Buell}}{{Gavil{\'a}n} et~al.}{2006}]{2006A&A...450..509G}
{Gavil{\'a}n} M.,  {Moll{\'a}} M.,    {Buell} J.~F.,  2006, \aap, 450, 509

\bibitem[\protect\citeauthoryear{{Guzman}, {Lucey} \& {Bower}}{{Guzman}
  et~al.}{1993}]{1993MNRAS.265..731G}
{Guzman} R.,  {Lucey} J.~R.,    {Bower} R.~G.,  1993, \mnras, 265, 731

\bibitem[\protect\citeauthoryear{{Harris}, {Harris}, {Layden} \&
  {Stetson}}{{Harris} et~al.}{2007}]{2007AJ....134...43H}
{Harris} W.~E.,  {Harris} G.~L.~H.,  {Layden} A.~C.,    {Stetson} P.~B.,  2007,
  \aj, 134, 43

\bibitem[\protect\citeauthoryear{{Iwamoto}, {Brachwitz}, {Nomoto}, {Kishimoto},
  {Umeda}, {Hix} \& {Thielemann}}{{Iwamoto} et~al.}{1999}]{1999ApJS..125..439I}
{Iwamoto} K.,  {Brachwitz} F.,  {Nomoto} K.,  {Kishimoto} N.,  {Umeda} H.,
  {Hix} W.~R.,    {Thielemann} F.-K.,  1999, \apjs, 125, 439

\bibitem[\protect\citeauthoryear{{Jungwiert}, {Carraro} \& {dalla
  Vecchia}}{{Jungwiert} et~al.}{2004}]{2004Ap&SS.289..441J}
{Jungwiert} B.,  {Carraro} G.,    {dalla Vecchia} C.,  2004, \apss, 289, 441

\bibitem[\protect\citeauthoryear{{Jungwiert}, {Combes} \& {Palou{\v
  s}}}{{Jungwiert} et~al.}{2001}]{2001A&A...376...85J}
{Jungwiert} B.,  {Combes} F.,    {Palou{\v s}} J.,  2001, \aap, 376, 85

\bibitem[\protect\citeauthoryear{{Katz}}{{Katz}}{1992}]{1992ApJ...391..502K}
{Katz} N.,  1992, \apj, 391, 502

\bibitem[\protect\citeauthoryear{{Kawata} \& {Gibson}}{{Kawata} \&
  {Gibson}}{2003}]{2003MNRAS.340..908K}
{Kawata} D.,  {Gibson} B.~K.,  2003, \mnras, 340, 908

\bibitem[\protect\citeauthoryear{{Kennicutt}
  Jr.}{{Kennicutt}}{1998}]{1998ApJ...498..541K}
{Kennicutt} Jr. R.~C.,  1998, \apj, 498, 541

\bibitem[\protect\citeauthoryear{{Klessen} \& {Lin}}{{Klessen} \&
  {Lin}}{2003}]{2003PhRvE..67d6311K}
{Klessen} R.~S.,  {Lin} D.~N.,  2003, \pre, 67, 046311

\bibitem[\protect\citeauthoryear{{Kobayashi}}{{Kobayashi}}{2004}]{2004MNRAS.34%
7..740K}
{Kobayashi} C.,  2004, \mnras, 347, 740

\bibitem[\protect\citeauthoryear{{Kobayashi}, {Springel} \&
  {White}}{{Kobayashi} et~al.}{2007}]{2007MNRAS.376.1465K}
{Kobayashi} C.,  {Springel} V.,    {White} S.~D.~M.,  2007, \mnras, 376, 1465

\bibitem[\protect\citeauthoryear{{Kroupa}}{{Kroupa}}{1998}]{1998MNRAS.298..231%
K}
{Kroupa} P.,  1998, \mnras, 298, 231

\bibitem[\protect\citeauthoryear{{Kuntschner}, {Ziegler}, {Sharples}, {Worthey}
  \& {Fricke}}{{Kuntschner} et~al.}{2002}]{2002A&A...395..761K}
{Kuntschner} H.,  {Ziegler} B.~L.,  {Sharples} R.~M.,  {Worthey} G.,
  {Fricke} K.~J.,  2002, \aap, 395, 761

\bibitem[\protect\citeauthoryear{{Lacey} \& {Fall}}{{Lacey} \&
  {Fall}}{1983}]{1983MNRAS.204..791L}
{Lacey} C.~G.,  {Fall} S.~M.,  1983, \mnras, 204, 791

\bibitem[\protect\citeauthoryear{{Lacey} \& {Fall}}{{Lacey} \&
  {Fall}}{1985}]{1985ApJ...290..154L}
{Lacey} C.~G.,  {Fall} S.~M.,  1985, ApJ, 290, 154

\bibitem[\protect\citeauthoryear{{Larsen}, {Brodie}, {Beasley} \&
  {Forbes}}{{Larsen} et~al.}{2002}]{2002AJ....124..828L}
{Larsen} S.~S.,  {Brodie} J.~P.,  {Beasley} M.~A.,    {Forbes} D.~A.,  2002,
  \aj, 124, 828

\bibitem[\protect\citeauthoryear{{Li}}{{Li}}{1991}]{1991JASA...86..316L}
{Li} K.~C.,  1991, Journ. Amer. Stat. Assoc., 86, 316

\bibitem[\protect\citeauthoryear{{Lia}, {Portinari} \& {Carraro}}{{Lia}
  et~al.}{2002}]{2002MNRAS.330..821L}
{Lia} C.,  {Portinari} L.,    {Carraro} G.,  2002, \mnras, 330, 821

\bibitem[\protect\citeauthoryear{{Lynden-Bell}}{{Lynden-Bell}}{1975}]{1975VA..%
...19..299L}
{Lynden-Bell} D.,  1975, Vistas in Astronomy, 19, 299

\bibitem[\protect\citeauthoryear{{Marigo}}{{Marigo}}{2001}]{2001A&A...370..194%
M}
{Marigo} P.,  2001, \aap, 370, 194

\bibitem[\protect\citeauthoryear{{Matteucci} \& {Francois}}{{Matteucci} \&
  {Francois}}{1989}]{1989MNRAS.239..885M}
{Matteucci} F.,  {Francois} P.,  1989, \mnras, 239, 885

\bibitem[\protect\citeauthoryear{{Matteucci} \& {Tornambe}}{{Matteucci} \&
  {Tornambe}}{1987}]{1987A&A...185...51M}
{Matteucci} F.,  {Tornambe} A.,  1987, \aap, 185, 51

\bibitem[\protect\citeauthoryear{{Merlin} \& {Chiosi}}{{Merlin} \&
  {Chiosi}}{2007}]{2007A&A...473..733M}
{Merlin} E.,  {Chiosi} C.,  2007, \aap, 473, 733

\bibitem[\protect\citeauthoryear{{Moll{\' a}} \& {D{\'{\i}}az}}{{Moll{\' a}} \&
  {D{\'{\i}}az}}{2005}]{2005MNRAS.358..521M}
{Moll{\' a}} M.,  {D{\'{\i}}az} A.~I.,  2005, MNRAS, 358, 521

\bibitem[\protect\citeauthoryear{{Molla}, {Ferrini} \& {Diaz}}{{Molla}
  et~al.}{1996}]{1996ApJ...466..668M}
{Molla} M.,  {Ferrini} F.,    {Diaz} A.~I.,  1996, \apj, 466, 668

\bibitem[\protect\citeauthoryear{{Moll{\'a}}, {V{\'{\i}}lchez}, {Gavil{\'a}n}
  \& {D{\'{\i}}az}}{{Moll{\'a}} et~al.}{2006}]{2006MNRAS.372.1069M}
{Moll{\'a}} M.,  {V{\'{\i}}lchez} J.~M.,  {Gavil{\'a}n} M.,    {D{\'{\i}}az}
  A.~I.,  2006, \mnras, 372, 1069

\bibitem[\protect\citeauthoryear{{Monaghan}}{{Monaghan}}{1992}]{1992ARA&A..30.%
.543M}
{Monaghan} J.~J.,  1992, \araa, 30, 543

\bibitem[\protect\citeauthoryear{{Monaghan}}{{Monaghan}}{2005}]{2005RPPh...68.%
1703M}
{Monaghan} J.~J.,  2005, Reports of Progress in Physics, 68, 1703

\bibitem[\protect\citeauthoryear{{Mosconi}, {Tissera}, {Lambas} \&
  {Cora}}{{Mosconi} et~al.}{2001}]{2001MNRAS.325...34M}
{Mosconi} M.~B.,  {Tissera} P.~B.,  {Lambas} D.~G.,    {Cora} S.~A.,  2001,
  \mnras, 325, 34

\bibitem[\protect\citeauthoryear{{O{\~n}orbe}, {Dom{\'{\i}}nguez-Tenreiro},
  {S{\'a}iz}, {Artal} \& {Serna}}{{O{\~n}orbe}
  et~al.}{2006}]{2006MNRAS.373..503O}
{O{\~n}orbe} J.,  {Dom{\'{\i}}nguez-Tenreiro} R.,  {S{\'a}iz} A.,  {Artal} H.,
    {Serna} A.,  2006, \mnras, 373, 503

\bibitem[\protect\citeauthoryear{{O{\~n}orbe}, {Dom{\'{\i}}nguez-Tenreiro},
  {S{\'a}iz} \& {Serna}}{{O{\~n}orbe} et~al.}{2007}]{2007MNRAS.376...39O}
{O{\~n}orbe} J.,  {Dom{\'{\i}}nguez-Tenreiro} R.,  {S{\'a}iz} A.,    {Serna}
  A.,  2007, \mnras, 376, 39

\bibitem[\protect\citeauthoryear{{O{\~n}orbe}, {Dom{\'{\i}}nguez-Tenreiro},
  {S{\'a}iz}, {Serna} \& {Artal}}{{O{\~n}orbe}
  et~al.}{2005}]{2005ApJ...632L..57O}
{O{\~n}orbe} J.,  {Dom{\'{\i}}nguez-Tenreiro} R.,  {S{\'a}iz} A.,  {Serna} A.,
    {Artal} H.,  2005, \apjl, 632, L57

\bibitem[\protect\citeauthoryear{{Pagel}}{{Pagel}}{1997}]{1997nceg.book.....P}
{Pagel} B.~E.~J.,  1997, {Nucleosynthesis and Chemical Evolution of Galaxies}.
Nucleosynthesis and Chemical Evolution of Galaxies, by Bernard E.~J.~Pagel,
  pp.~392.~ISBN 0521550610.~Cambridge, UK: Cambridge University Press, October
  1997.

\bibitem[\protect\citeauthoryear{{Pagel} \& {Patchett}}{{Pagel} \&
  {Patchett}}{1975}]{1975MNRAS.172...13P}
{Pagel} B.~E.~J.,  {Patchett} B.~E.,  1975, \mnras, 172, 13

\bibitem[\protect\citeauthoryear{{Persic}, {Salucci} \& {Stel}}{{Persic}
  et~al.}{1996}]{1996MNRAS.281...27P}
{Persic} M.,  {Salucci} P.,    {Stel} F.,  1996, \mnras, 281, 27

\bibitem[\protect\citeauthoryear{{Portinari}, {Chiosi} \&
  {Bressan}}{{Portinari} et~al.}{1998}]{1998A&A...334..505P}
{Portinari} L.,  {Chiosi} C.,    {Bressan} A.,  1998, \aap, 334, 505

\bibitem[\protect\citeauthoryear{{Puzia}, {Kissler-Patig}, {Thomas},
  {Maraston}, {Saglia}, {Bender}, {Goudfrooij} \& {Hempel}}{{Puzia}
  et~al.}{2005}]{2005A&A...439..997P}
{Puzia} T.~H.,  {Kissler-Patig} M.,  {Thomas} D.,  {Maraston} C.,  {Saglia}
  R.~P.,  {Bender} R.,  {Goudfrooij} P.,    {Hempel} M.,  2005, \aap, 439, 997

\bibitem[\protect\citeauthoryear{{Raiteri}, {Villata} \& {Navarro}}{{Raiteri}
  et~al.}{1996}]{1996A&A...315..105R}
{Raiteri} C.~M.,  {Villata} M.,    {Navarro} J.~F.,  1996, \aap, 315, 105

\bibitem[\protect\citeauthoryear{{Reetz}}{{Reetz}}{1999}]{1999Ap&SS.265..171R}
{Reetz} J.,  1999, \apss, 265, 171

\bibitem[\protect\citeauthoryear{{Romeo}, {Portinari} \&
  {Sommer-Larsen}}{{Romeo} et~al.}{2005}]{2005MNRAS.361..983R}
{Romeo} A.~D.,  {Portinari} L.,    {Sommer-Larsen} J.,  2005, \mnras, 361, 983

\bibitem[\protect\citeauthoryear{{Romeo}, {Sommer-Larsen}, {Portinari} \&
  {Antonuccio-Delogu}}{{Romeo} et~al.}{2006}]{2006MNRAS.371..548R}
{Romeo} A.~D.,  {Sommer-Larsen} J.,  {Portinari} L.,    {Antonuccio-Delogu} V.,
   2006, \mnras, 371, 548

\bibitem[\protect\citeauthoryear{{Ruiz-Lapuente}, {Blinnikov}, {Canal},
  {Mendez}, {Sorokina}, {Visco} \& {Walton}}{{Ruiz-Lapuente}
  et~al.}{2000}]{2000MmSAI..71..435R}
{Ruiz-Lapuente} P.,  {Blinnikov} S.,  {Canal} R.,  {Mendez} J.,  {Sorokina} E.,
   {Visco} A.,    {Walton} N.,  2000, Memorie della Societa Astronomica
  Italiana, 71, 435

\bibitem[\protect\citeauthoryear{{S{\'a}iz}, {Dom{\'{\i}}nguez-Tenreiro} \&
  {Serna}}{{S{\'a}iz} et~al.}{2002}]{2002Ap&SS.281..309S}
{S{\'a}iz} A.,  {Dom{\'{\i}}nguez-Tenreiro} R.,    {Serna} A.,  2002, \apss,
  281, 309

\bibitem[\protect\citeauthoryear{{Salpeter}}{{Salpeter}}{1955}]{1955ApJ...121.%
.161S}
{Salpeter} E.~E.,  1955, \apj, 121, 161

\bibitem[\protect\citeauthoryear{{Sansom}, {Peace} \& {Dodd}}{{Sansom}
  et~al.}{1994}]{1994MNRAS.271...39S}
{Sansom} A.~E.,  {Peace} G.,    {Dodd} M.,  1994, \mnras, 271, 39

\bibitem[\protect\citeauthoryear{{Scannapieco}, {Tissera}, {White} \&
  {Springel}}{{Scannapieco} et~al.}{2005}]{2005MNRAS.364..552S}
{Scannapieco} C.,  {Tissera} P.~B.,  {White} S.~D.~M.,    {Springel} V.,  2005,
  \mnras, 364, 552

\bibitem[\protect\citeauthoryear{{Schaller}, {Schaerer}, {Meynet} \&
  {Maeder}}{{Schaller} et~al.}{1992}]{1992A&AS...96..269S}
{Schaller} G.,  {Schaerer} D.,  {Meynet} G.,    {Maeder} A.,  1992, \aaps, 96,
  269

\bibitem[\protect\citeauthoryear{{Scott}}{{Scott}}{1992}]{1992mde..book.....S}
{Scott} D.~W.,  1992, {Multivariate Density Estimation}.
Multivariate Density Estimation, Wiley, New York, 1992

\bibitem[\protect\citeauthoryear{{Serna}, {Dom{\'i}nguez-Tenreiro} \&
  {S{\'a}iz}}{{Serna} et~al.}{2003}]{2003ApJ...597..878S}
{Serna} A.,  {Dom{\'i}nguez-Tenreiro} R.,    {S{\'a}iz} A.,  2003, \apj, 597,
  878

\bibitem[\protect\citeauthoryear{{Silk}}{{Silk}}{2001}]{2001MNRAS.324..313S}
{Silk} J.,  2001, \mnras, 324, 313

\bibitem[\protect\citeauthoryear{{Sommer-Larsen}, {Romeo} \&
  {Portinari}}{{Sommer-Larsen} et~al.}{2005}]{2005MNRAS.357..478S}
{Sommer-Larsen} J.,  {Romeo} A.~D.,    {Portinari} L.,  2005, \mnras, 357, 478

\bibitem[\protect\citeauthoryear{{Sommer-Larsen} \& {Yoshii}}{{Sommer-Larsen}
  \& {Yoshii}}{1989}]{1989MNRAS.238..133S}
{Sommer-Larsen} J.,  {Yoshii} Y.,  1989, MNRAS, 238, 133

\bibitem[\protect\citeauthoryear{{Springel} \& {Hernquist}}{{Springel} \&
  {Hernquist}}{2003}]{2003MNRAS.339..289S}
{Springel} V.,  {Hernquist} L.,  2003, \mnras, 339, 289

\bibitem[\protect\citeauthoryear{{Steinmetz} \& {Muller}}{{Steinmetz} \&
  {Muller}}{1994}]{1994A&A...281L..97S}
{Steinmetz} M.,  {Muller} E.,  1994, \aap, 281, L97

\bibitem[\protect\citeauthoryear{{Steinmetz} \& {Muller}}{{Steinmetz} \&
  {Muller}}{1995}]{1995MNRAS.276..549S}
{Steinmetz} M.,  {Muller} E.,  1995, \mnras, 276, 549

\bibitem[\protect\citeauthoryear{{Sutherland} \& {Dopita}}{{Sutherland} \&
  {Dopita}}{1993}]{1993ApJS...88..253S}
{Sutherland} R.~S.,  {Dopita} M.~A.,  1993, \apjs, 88, 253

\bibitem[\protect\citeauthoryear{{Talbot} Jr. \& {Arnett}}{{Talbot} \&
  {Arnett}}{1973}]{1973ApJ...186...51T}
{Talbot} Jr. R.~J.,  {Arnett} W.~D.,  1973, \apj, 186, 51

\bibitem[\protect\citeauthoryear{{Taylor}}{{Taylor}}{1921}]{taylor:21}
{Taylor} G.,  1921, Proc. Lond. Math. Soc, 20, 196

\bibitem[\protect\citeauthoryear{{Thomas}, {Maraston}, {Bender} \& {Mendes de
  Oliveira}}{{Thomas} et~al.}{2005}]{2005ApJ...621..673T}
{Thomas} D.,  {Maraston} C.,  {Bender} R.,    {Mendes de Oliveira} C.,  2005,
  \apj, 621, 673

\bibitem[\protect\citeauthoryear{{Timmes}, {Woosley} \& {Weaver}}{{Timmes}
  et~al.}{1995}]{1995ApJS...98..617T}
{Timmes} F.~X.,  {Woosley} S.~E.,    {Weaver} T.~A.,  1995, \apjs, 98, 617

\bibitem[\protect\citeauthoryear{{Tinsley}}{{Tinsley}}{1980}]{1980FCPh....5..2%
87T}
{Tinsley} B.~M.,  1980, Fundamentals of Cosmic Physics, 5, 287

\bibitem[\protect\citeauthoryear{{Valle}, {Ferrini}, {Galli} \&
  {Shore}}{{Valle} et~al.}{2002}]{2002ApJ...566..252V}
{Valle} G.,  {Ferrini} F.,  {Galli} D.,    {Shore} S.~N.,  2002, ApJ, 566, 252

\bibitem[\protect\citeauthoryear{{van den Hoek} \& {Groenewegen}}{{van den
  Hoek} \& {Groenewegen}}{1997}]{1997A&AS..123..305V}
{van den Hoek} L.~B.,  {Groenewegen} M.~A.~T.,  1997, A\&A Supl.Series, 123,
  305

\bibitem[\protect\citeauthoryear{{Weisberg}}{{Weisberg}}{2002}]{2002JSS.....7.%
...0W}
{Weisberg} S.,  2002, Journ. Statistical Software, 7

\bibitem[\protect\citeauthoryear{{Woosley} \& {Weaver}}{{Woosley} \&
  {Weaver}}{1995}]{1995ApJS..101..181W}
{Woosley} S.~E.,  {Weaver} T.~A.,  1995, \apjs, 101, 181

\bibitem[\protect\citeauthoryear{{Worthey}}{{Worthey}}{1994}]{1994ApJS...95..1%
07W}
{Worthey} G.,  1994, \apjs, 95, 107

\bibitem[\protect\citeauthoryear{{Worthey}, {Faber} \& {Gonzalez}}{{Worthey}
  et~al.}{1992}]{1992ApJ...398...69W}
{Worthey} G.,  {Faber} S.~M.,    {Gonzalez} J.~J.,  1992, \apj, 398, 69

\bibitem[\protect\citeauthoryear{{Yepes}, {Kates}, {Khokhlov} \&
  {Klypin}}{{Yepes} et~al.}{1997}]{1997MNRAS.284..235Y}
{Yepes} G.,  {Kates} R.,  {Khokhlov} A.,    {Klypin} A.,  1997, \mnras, 284,
  235

\end{thebibliography}

\renewcommand{\theequation}{A-\arabic{equation}}
\setcounter{equation}{0}  
\section*{Appendix A: Analytical MDF for a closed box model without instantaneous recycling} \label{appendixa}

Following \citet{1997nceg.book.....P}, we consider a closed box filled with primordial gas. We assume a \citet{2003PASP..115..763C} IMF, $\Phi(M)$, and the star-forming law $\Psi(t)$ given by Eq. (\ref{eq:expsfr}). We denote the gas and star fractions by $g(t)$ and $s(t)$, respectively. The mass conservation then implies that
\begin{equation}
 g(t)+s(t)=1\;,
\end{equation}
and therefore
\begin{equation}\label{eq:dgds}
 \frac{\mathrm{d}s}{\mathrm{d}t}=-\frac{\mathrm{d}g}{\mathrm{d}t}=\Psi(t)-e(t)\;, 
\end{equation}
where $e(t)$ is the ejection rate of gas:
\begin{equation}
 e(t)=\int_{M_\tau(t)}^{M_u} (M-M_r(M)) \Psi(t - \tau(M)) \frac{\Phi(M)}{M} \mathrm{d}M\;.
\end{equation}

Here $\tau(M)$ and $M_\tau(t)$ represent the lifetime of a star of mass $M$ and its inverse, respectively.

Finding the MDF is equivalent to expressing the star formation rate as a function of the metallicity. We must then obtain the time dependence of the gas composition, $Z(t)$, and invert such a function to find $t(Z)$. By substituting the latter function into the star formation rate $\Psi(t)$, we will have $\Psi(Z)$.

To find $Z(t)$, we notice that the time derivative of the total content of metals $g(t)Z(t)$ must be equal to the total ejections by dying stars $e_Z(t)$ minus the mass of metals that gets trapped in newly formed stars $\Psi(t)Z(t)$:
\begin{equation}
 \frac{\mathrm{d}(Zg)}{\mathrm{d}t}=e_Z(t)-\Psi(t)Z(t)\;.
\end{equation}

Using Eq. \ref{eq:dgds} we can simplify this expression:
\begin{equation}
 g(t)\frac{\mathrm{d}Z}{\mathrm{d}t}=e_Z(t)-e(t)Z(t)\;.
\end{equation}

The expression for $e_Z(t)$ is similar to that of $e(t)$:
\begin{eqnarray}
 e_Z(t) & = & \int_{M_\tau(t)}^{M_u} \{ \left[ M-M_r(M)\right] Z(t - \tau(M)) + M\,q_Z(M) \} \nonumber\\
 & & \Psi(t - \tau(M)) \frac{\Phi(M)}{M} \mathrm{d}M\;,
\end{eqnarray}
where $q_Z(M)$ is in our case the oxygen yield for a star of mass $M$. This equation involves $Z(t)$ inside the integral, so we end up with two integro-differential coupled equations. To solve them we run an iterative procedure assuming an initial value for $Z(t)$, namely $Z(t)=0$, obtaining an expression for $e_Z(t)$, and recomputing the value of $Z(t)$ until it converges. Since both $M_r(M)$ and $q_Z(M)$ are specified in the yield tables described in section \ref{sec:metalprod} and do not have a simple analytical expression, all the procedure has to be run numerically, using interpolating functions for both functions and a suitable integrator.

Once $Z(t)$ and its inverse $t_Z(Z)$ are known, we compute the final MDF as
\begin{eqnarray}
 \mathrm{MDF}(\mathrm{log}_{10}(Z)) & = & \Psi(t_Z(Z)) \left[1-E(t_1-t_Z(Z))\right]\nonumber \\
 & & \frac{\mathrm{d}t_Z(Z)}{\mathrm{d}Z} \frac{\mathrm{d}Z}{\mathrm{d}(\mathrm{log}_{10}(Z))}\;.
\end{eqnarray}

This is simply the aforementioned star formation expressed as a function of metallicity $\Psi(t_Z(Z))$, with suitable variable exchange terms, and a correction term $\left[1-E(t_1-t_Z(Z))\right]$ similar to that already introduced in Eqs. (\ref{eq:probability}, \ref{eq:spread}). This term accounts for the mass fraction of stars formed with a given metallicity that died before the final time limit $t_1$ (i.e. the most massive ones).


\begin{table*}
 \label{tab:c_T}
 \caption{DRR $\mathbf{c}=(c_{He}, c_C,..., c_{Fe})$ coefficients as a function of temperature}
 \begin{tabular}{l r r r r r r r r r r r}
\hline
log$_{10}$(T)&
C&
O&
N&
Ne&
Mg&
Si&
S&
Ca&
Fe&
$\zeta_0$&
$\zeta_1$\\
\hline
4.0 & 2.23e-02 & 4.53e-01 & 6.13e-02 & -4.55e-02 & 3.30e-02 & 1.58e-01 & -8.91e-02 & 1.08e-01 & 5.33e-01 & 6.66e-06 & 3.80e+01\\
4.1 & 1.45e-02 & 1.47e-03 & -1.31e-02 & -2.80e-03 & 1.21e-01 & -7.55e-02 & 3.04e-01 & -9.40e-01 & 3.91e-02 & 4.46e-07 & 1.57e+03\\
4.2 & 2.61e-03 & -2.93e-03 & -1.15e-02 & 4.87e-05 & 8.46e-02 & -6.21e-02 & 2.33e-01 & -9.66e-01 & 8.49e-03 & 1.98e-08 & 8.69e+03\\
4.3 & 1.87e-02 & 7.87e-03 & -4.55e-03 & -5.84e-03 & 4.49e-02 & -6.73e-02 & 2.68e-01 & -9.58e-01 & 1.43e-02 & 3.64e-07 & 1.51e+03\\
4.4 & 2.12e-01 & 8.39e-02 & 6.78e-02 & -3.14e-02 & -1.87e-01 & 1.39e-01 & 2.16e-01 & -9.13e-01 & 7.61e-02 & 4.28e-06 & 1.70e+02\\
4.5 & 9.08e-02 & 3.65e-02 & 4.01e-02 & -1.34e-02 & -1.59e-01 & 1.34e-01 & -2.36e-01 & 9.41e-01 & 2.35e-02 & 1.83e-06 & 4.46e+02\\
4.6 & 6.87e-02 & 3.10e-02 & 3.59e-02 & -1.06e-02 & -1.37e-01 & 1.12e-01 & -2.35e-01 & 9.51e-01 & 1.87e-02 & 1.42e-06 & 5.50e+02\\
4.7 & 4.13e-02 & 2.45e-02 & 2.91e-02 & -8.16e-03 & -1.12e-01 & 9.77e-02 & -2.40e-01 & 9.57e-01 & 1.16e-02 & 9.30e-07 & 7.58e+02\\
4.8 & 3.47e-02 & 2.48e-02 & 3.32e-02 & -6.86e-03 & -1.02e-01 & 8.68e-02 & -2.30e-01 & 9.61e-01 & 1.18e-02 & 8.51e-07 & 7.50e+02\\
4.9 & 3.33e-02 & 2.58e-02 & 3.81e-02 & -6.08e-03 & -1.09e-01 & 7.53e-02 & -2.34e-01 & 9.60e-01 & 1.13e-02 & 8.32e-07 & 7.54e+02\\
5.0 & 2.40e-02 & 2.41e-02 & 3.52e-02 & -4.16e-03 & -1.03e-01 & 6.41e-02 & -2.25e-01 & 9.65e-01 & 9.18e-03 & 6.67e-07 & 8.37e+02\\
5.1 & 5.90e-03 & 2.59e-02 & 3.44e-02 & -3.44e-03 & -1.13e-01 & 5.59e-02 & -2.12e-01 & 9.68e-01 & 7.72e-03 & 4.11e-07 & 8.65e+02\\
5.2 & 4.36e-04 & 3.16e-02 & 3.60e-02 & -1.79e-03 & -1.22e-01 & 6.08e-02 & -2.21e-01 & 9.64e-01 & 8.87e-03 & 3.79e-07 & 6.90e+02\\
5.3 & -2.21e-03 & 3.71e-02 & 2.40e-02 & 3.62e-03 & -1.09e-01 & 7.86e-02 & -2.56e-01 & 9.54e-01 & 7.16e-03 & 3.26e-07 & 5.36e+02\\
5.4 & -2.14e-03 & 3.84e-02 & 1.34e-02 & 9.12e-03 & -9.65e-02 & 8.33e-02 & -2.71e-01 & 9.52e-01 & 7.75e-03 & 3.10e-07 & 4.83e+02\\
5.5 & -1.21e-03 & 2.47e-02 & 9.45e-03 & 2.16e-02 & -7.99e-02 & 8.94e-02 & -2.72e-01 & 9.53e-01 & 4.32e-03 & 2.37e-07 & 6.21e+02\\
5.6 & -9.93e-04 & 1.07e-02 & 6.16e-03 & 2.91e-02 & -6.56e-02 & 7.52e-02 & -2.40e-01 & 9.65e-01 & 2.31e-03 & 1.55e-07 & 9.87e+02\\
5.7 & -6.68e-04 & 7.54e-03 & 5.72e-03 & 4.18e-02 & -5.83e-02 & 7.71e-02 & -2.35e-01 & 9.66e-01 & 3.69e-03 & 1.72e-07 & 8.87e+02\\
5.8 & -1.05e-03 & 7.79e-03 & 8.13e-03 & 4.74e-02 & -5.31e-02 & 8.66e-02 & -2.37e-01 & 9.65e-01 & 1.11e-02 & 2.45e-07 & 7.64e+02\\
5.9 & -1.64e-03 & 1.02e-02 & 7.11e-03 & 2.46e-02 & -1.56e-02 & 1.35e-01 & -2.57e-01 & 9.54e-01 & 4.40e-02 & 5.10e-07 & 7.61e+02\\
6.0 & 3.08e-03 & 9.11e-03 & 9.36e-03 & 1.30e-02 & 6.56e-02 & 2.62e-01 & -2.68e-01 & 8.75e-01 & 1.55e-01 & 1.47e-06 & 5.16e+02\\
6.1 & 3.64e-03 & 8.29e-03 & 1.24e-02 & 8.67e-04 & -2.71e-02 & 2.08e-01 & -1.76e-01 & 8.88e-01 & 1.65e-01 & 1.49e-06 & 7.27e+02\\
6.2 & 3.72e-03 & 1.08e-02 & 1.56e-02 & 4.35e-04 & -4.39e-02 & 2.79e-01 & -1.04e-01 & 7.83e-01 & 2.83e-01 & 2.41e-06 & 4.88e+02\\
6.3 & 3.51e-03 & 1.02e-02 & 1.17e-02 & -1.11e-03 & -4.37e-02 & 2.33e-01 & -1.44e-01 & 9.00e-01 & 1.57e-01 & 1.49e-06 & 6.54e+02\\
6.4 & 1.88e-03 & 1.52e-02 & 8.75e-03 & -9.09e-04 & -8.42e-02 & 1.96e-01 & -1.29e-01 & 9.47e-01 & 1.04e-01 & 1.09e-06 & 6.76e+02\\
6.5 & 1.13e-03 & 1.81e-02 & 1.31e-02 & 1.10e-02 & -6.55e-02 & 2.18e-01 & -1.28e-01 & 9.03e-01 & 1.87e-01 & 1.70e-06 & 4.63e+02\\
6.6 & 1.34e-03 & 1.21e-02 & 8.44e-03 & 1.32e-02 & -4.33e-02 & 1.22e-01 & -1.35e-01 & 9.53e-01 & 1.74e-01 & 1.45e-06 & 6.46e+02\\
6.7 & 2.60e-03 & 1.27e-02 & 1.27e-02 & 2.20e-02 & -2.92e-02 & 1.40e-01 & -1.52e-01 & 9.07e-01 & 2.53e-01 & 2.03e-06 & 5.04e+02\\
6.8 & 4.29e-03 & 1.65e-02 & 2.09e-02 & 2.78e-02 & -4.33e-03 & 1.83e-01 & -1.48e-01 & 7.94e-01 & 3.95e-01 & 3.10e-06 & 3.56e+02\\
6.9 & 3.97e-03 & 1.52e-02 & 2.42e-02 & 2.66e-02 & 2.76e-02 & 1.94e-01 & -9.23e-02 & 6.20e-01 & 5.66e-01 & 4.27e-06 & 3.09e+02\\
7.0 & 1.18e-02 & 1.27e-02 & 2.76e-02 & 2.47e-02 & 6.48e-02 & 1.71e-01 & 4.97e-02 & 2.48e-03 & 7.01e-01 & 5.26e-06 & 2.84e+02\\
7.1 & 1.06e-02 & 1.85e-02 & 2.92e-02 & 2.22e-02 & 1.49e-02 & 1.85e-01 & -3.01e-04 & 3.97e-01 & 6.81e-01 & 5.16e-06 & 2.74e+02\\
7.2 & 7.80e-03 & 2.34e-02 & 2.80e-02 & 1.88e-02 & -1.84e-02 & 1.73e-01 & -4.31e-02 & 6.52e-01 & 5.47e-01 & 4.24e-06 & 2.92e+02\\
7.3 & 7.61e-03 & 2.15e-02 & 1.98e-02 & 2.02e-02 & -6.51e-03 & 1.42e-01 & -4.50e-02 & 7.87e-01 & 4.14e-01 & 3.29e-06 & 3.30e+02\\
7.4 & 9.91e-03 & 2.80e-02 & 1.80e-02 & 2.13e-02 & -3.79e-02 & 1.37e-01 & -5.00e-02 & 8.14e-01 & 4.01e-01 & 3.26e-06 & 3.04e+02\\
7.5 & 1.37e-02 & 3.87e-02 & 1.52e-02 & 3.14e-02 & -3.30e-02 & 1.76e-01 & -4.85e-02 & 7.47e-01 & 5.04e-01 & 4.11e-06 & 2.24e+02\\
7.6 & 1.13e-02 & 3.94e-02 & 1.08e-02 & 2.86e-02 & -6.82e-02 & 1.77e-01 & -1.24e-01 & 8.45e-01 & 4.34e-01 & 3.56e-06 & 2.45e+02\\
7.7 & 1.50e-02 & 6.76e-02 & 1.28e-02 & 5.85e-02 & -5.65e-02 & 2.77e-01 & -5.40e-02 & 5.18e-01 & 7.45e-01 & 6.02e-06 & 1.36e+02\\
7.8 & 2.16e-02 & 7.98e-02 & 7.41e-03 & 7.25e-02 & -2.17e-03 & 2.99e-01 & 3.78e-02 & -4.54e-02 & 8.52e-01 & 6.93e-06 & 1.13e+02\\
7.9 & 2.00e-02 & 6.52e-02 & 4.17e-03 & 5.68e-02 & -3.31e-02 & 2.70e-01 & -1.34e-01 & 6.26e-01 & 6.21e-01 & 5.19e-06 & 1.45e+02\\
8.0 & 1.98e-02 & 6.66e-02 & -1.28e-03 & 6.68e-02 & 4.44e-02 & 2.20e-01 & 1.16e-01 & -5.75e-01 & 6.37e-01 & 5.26e-06 & 1.34e+02\\
8.1 & 1.54e-02 & 5.59e-02 & 1.15e-03 & 4.29e-02 & -3.62e-03 & 9.26e-02 & 2.16e-01 & -8.28e-01 & 4.37e-01 & 3.65e-06 & 1.82e+02\\
8.2 & 1.61e-02 & 6.35e-02 & -6.72e-04 & 3.80e-02 & -1.82e-02 & 3.22e-02 & 2.94e-01 & -8.25e-01 & 4.34e-01 & 3.63e-06 & 1.75e+02\\
8.3 & 1.00e-02 & 3.93e-02 & -3.32e-03 & 2.82e-02 & 4.48e-03 & 8.93e-03 & 2.60e-01 & -9.18e-01 & 2.69e-01 & 2.24e-06 & 2.71e+02\\
8.4 & 7.80e-03 & 2.82e-02 & -7.31e-03 & 2.38e-02 & 2.61e-02 & -3.57e-03 & 2.47e-01 & -9.46e-01 & 1.98e-01 & 1.64e-06 & 3.56e+02\\
8.5 & 6.08e-03 & 2.13e-02 & -7.85e-03 & 2.08e-02 & 4.32e-02 & -1.34e-02 & 2.41e-01 & -9.56e-01 & 1.57e-01 & 1.28e-06 & 4.41e+02\\
\hline
\end{tabular}
\end{table*}

\begin{table*}
 \label{tab:Lambda}
 \caption{Cooling function log$ _{10}(\Lambda_N)$ as a function of temperature T and the metallicity parameter $\zeta$}
 \begin{tabular}{l r r r r r r r r r r r}
\hline
log$_{10}$(T)&
$\zeta$=0&
$\zeta$=0.01&
$\zeta$=0.02&
$\zeta$=0.03&
$\zeta$=0.04&
$\zeta$=0.1&
$\zeta$=0.2&
$\zeta$=0.4&
$\zeta$=0.6&
$\zeta$=0.8&
$\zeta$=1\\
\hline
4.0 & 21.22 &  21.50 &  21.66 &  21.78 &  21.88 &  22.21 &  22.48 &  22.75 &  22.90 &  22.99 & 23.07\\
4.1 & 23.80 &  23.80 &  23.80 &  23.81 &  23.81 &  23.84 &  23.88 &  23.95 &  24.00 &  24.04 & 24.08\\
4.2 & 25.21 &  25.21 &  25.21 &  25.21 &  25.21 &  25.22 &  25.23 &  25.25 &  25.26 &  25.28 & 25.29\\
4.3 & 25.30 &  25.30 &  25.31 &  25.31 &  25.32 &  25.35 &  25.39 &  25.45 &  25.50 &  25.54 & 25.57\\
4.4 & 25.09 &  25.11 &  25.13 &  25.14 &  25.16 &  25.24 &  25.35 &  25.50 &  25.60 &  25.67 & 25.72\\
4.5 & 24.93 &  24.98 &  25.03 &  25.07 &  25.10 &  25.27 &  25.46 &  25.67 &  25.81 &  25.90 & 25.97\\
4.6 & 24.77 &  24.89 &  24.98 &  25.06 &  25.12 &  25.38 &  25.62 &  25.88 &  26.03 &  26.13 & 26.21\\
4.7 & 24.70 &  24.90 &  25.04 &  25.14 &  25.23 &  25.54 &  25.80 &  26.07 &  26.22 &  26.33 & 26.41\\
4.8 & 24.97 &  25.13 &  25.25 &  25.35 &  25.43 &  25.72 &  25.98 &  26.24 &  26.40 &  26.50 & 26.58\\
4.9 & 25.24 &  25.37 &  25.47 &  25.55 &  25.62 &  25.89 &  26.13 &  26.39 &  26.55 &  26.65 & 26.73\\
5.0 & 25.16 &  25.35 &  25.48 &  25.59 &  25.67 &  25.98 &  26.24 &  26.51 &  26.67 &  26.78 & 26.86\\
5.1 & 24.94 &  25.27 &  25.46 &  25.59 &  25.69 &  26.04 &  26.32 &  26.60 &  26.76 &  26.87 & 26.96\\
5.2 & 24.71 &  25.25 &  25.48 &  25.63 &  25.75 &  26.12 &  26.41 &  26.69 &  26.86 &  26.97 & 27.06\\
5.3 & 24.52 &  25.25 &  25.51 &  25.67 &  25.79 &  26.18 &  26.47 &  26.75 &  26.92 &  27.03 & 27.12\\
5.4 & 24.27 &  25.21 &  25.49 &  25.66 &  25.78 &  26.17 &  26.46 &  26.75 &  26.91 &  27.02 & 27.11\\
5.5 & 24.27 &  24.97 &  25.22 &  25.38 &  25.50 &  25.88 &  26.17 &  26.46 &  26.62 &  26.74 & 26.83\\
5.6 & 24.20 &  24.77 &  25.01 &  25.16 &  25.27 &  25.65 &  25.94 &  26.22 &  26.39 &  26.50 & 26.59\\
5.7 & 24.14 &  24.72 &  24.97 &  25.12 &  25.24 &  25.62 &  25.91 &  26.19 &  26.36 &  26.47 & 26.56\\
5.8 & 24.09 &  24.64 &  24.88 &  25.03 &  25.15 &  25.52 &  25.81 &  26.10 &  26.26 &  26.38 & 26.47\\
5.9 & 24.12 &  24.43 &  24.61 &  24.74 &  24.84 &  25.18 &  25.46 &  25.74 &  25.90 &  26.02 & 26.11\\
6.0 & 24.12 &  24.34 &  24.48 &  24.59 &  24.68 &  24.99 &  25.26 &  25.54 &  25.70 &  25.82 & 25.90\\
6.1 & 24.12 &  24.30 &  24.43 &  24.53 &  24.61 &  24.92 &  25.18 &  25.45 &  25.62 &  25.73 & 25.82\\
6.2 & 24.12 &  24.30 &  24.42 &  24.52 &  24.60 &  24.90 &  25.16 &  25.44 &  25.60 &  25.71 & 25.80\\
6.3 & 24.12 &  24.27 &  24.38 &  24.46 &  24.54 &  24.82 &  25.07 &  25.34 &  25.50 &  25.62 & 25.70\\
6.4 & 24.13 &  24.22 &  24.30 &  24.36 &  24.42 &  24.66 &  24.89 &  25.15 &  25.30 &  25.41 & 25.50\\
6.5 & 24.15 &  24.21 &  24.26 &  24.31 &  24.35 &  24.54 &  24.74 &  24.97 &  25.12 &  25.23 & 25.30\\
6.6 & 24.17 &  24.22 &  24.25 &  24.29 &  24.32 &  24.48 &  24.66 &  24.88 &  25.02 &  25.12 & 25.20\\
6.7 & 24.20 &  24.23 &  24.26 &  24.29 &  24.32 &  24.45 &  24.60 &  24.80 &  24.94 &  25.03 & 25.11\\
6.8 & 24.23 &  24.26 &  24.28 &  24.30 &  24.32 &  24.43 &  24.56 &  24.75 &  24.87 &  24.96 & 25.03\\
6.9 & 24.27 &  24.29 &  24.31 &  24.33 &  24.34 &  24.44 &  24.56 &  24.73 &  24.84 &  24.93 & 25.00\\
7.0 & 24.30 &  24.32 &  24.34 &  24.35 &  24.37 &  24.45 &  24.57 &  24.72 &  24.84 &  24.92 & 24.99\\
7.1 & 24.34 &  24.35 &  24.37 &  24.38 &  24.39 &  24.46 &  24.55 &  24.69 &  24.80 &  24.88 & 24.94\\
7.2 & 24.38 &  24.39 &  24.40 &  24.41 &  24.42 &  24.47 &  24.54 &  24.66 &  24.75 &  24.82 & 24.87\\
7.3 & 24.42 &  24.42 &  24.43 &  24.44 &  24.45 &  24.49 &  24.54 &  24.64 &  24.71 &  24.78 & 24.83\\
7.4 & 24.46 &  24.46 &  24.47 &  24.48 &  24.48 &  24.51 &  24.56 &  24.64 &  24.71 &  24.76 & 24.81\\
7.5 & 24.50 &  24.51 &  24.51 &  24.52 &  24.52 &  24.55 &  24.59 &  24.66 &  24.71 &  24.76 & 24.80\\
7.6 & 24.54 &  24.55 &  24.55 &  24.56 &  24.56 &  24.58 &  24.62 &  24.68 &  24.73 &  24.77 & 24.81\\
7.7 & 24.59 &  24.59 &  24.60 &  24.60 &  24.60 &  24.62 &  24.65 &  24.71 &  24.76 &  24.79 & 24.83\\
7.8 & 24.63 &  24.64 &  24.64 &  24.64 &  24.65 &  24.66 &  24.69 &  24.74 &  24.78 &  24.82 & 24.86\\
7.9 & 24.68 &  24.68 &  24.68 &  24.69 &  24.69 &  24.71 &  24.73 &  24.78 &  24.82 &  24.85 & 24.88\\
8.0 & 24.72 &  24.73 &  24.73 &  24.73 &  24.74 &  24.75 &  24.77 &  24.82 &  24.85 &  24.88 & 24.91\\
8.1 & 24.77 &  24.77 &  24.78 &  24.78 &  24.78 &  24.79 &  24.82 &  24.85 &  24.89 &  24.92 & 24.95\\
8.2 & 24.82 &  24.82 &  24.82 &  24.82 &  24.83 &  24.84 &  24.86 &  24.90 &  24.93 &  24.96 & 24.98\\
8.3 & 24.87 &  24.87 &  24.87 &  24.87 &  24.87 &  24.89 &  24.90 &  24.94 &  24.97 &  24.99 & 25.02\\
8.4 & 24.91 &  24.92 &  24.92 &  24.92 &  24.92 &  24.93 &  24.95 &  24.98 &  25.01 &  25.03 & 25.06\\
8.5 & 24.96 &  24.96 &  24.96 &  24.97 &  24.97 &  24.98 &  25.00 &  25.03 &  25.05 &  25.08 & 25.10\\
\hline
\end{tabular}
\end{table*}

\label{lastpage}
\end{document}